# Machine Learning-Driven Content Popularity Prediction and Cache Optimization in D2D Clustered Networks

Maede Rezaei, Ahmadreza Montazerolghaem

**Abstract**—Advancements in wireless communication technology have led to the widespread use of smart devices—including computers, mobile phones, tablets, wearable devices, and vehicles—which has significantly increased the demand for high-quality content. This growing demand puts pressure on the backhaul links in cellular networks, resulting in congestion and content delivery delays. To address this, cache-enabled networks and edge caching, such as caching in user devices, have emerged as promising solutions to reduce backhaul traffic. By caching content locally and using device-to-device (D2D) communication for retrieval, content delivery can be made more efficient. However, limited cache capacity requires intelligent content selection strategies. The popularity of the content is dynamic and varies with user preferences, where less than 20% of the users generate 80% of multimedia traffic. Many existing methods fail to consider this user heterogeneity, often assuming uniform preferences throughout the network. This paper proposes a novel *Machine Learning-Driven Content Popularity Prediction* and *Cache Optimization* (ML-CPCO) framework that dynamically predicts user- and cluster-level content demand, incorporates user willingness to participate in caching, and optimizes cache placement in *D2D-enabled clustered networks*. The system predicts future content requests using machine learning algorithms and estimates content popularity at the cluster level. Based on these predictions, cache placement decisions are made to maximize efficiency. The simulation results show that the proposed approach performs well under various network conditions, achieving a cache utilization rate of nearly 97%—the highest among the methods compared. In addition, it offers an improved hit rate with an acceptable execution time, resulting in reduced backhaul traffic and enhanced user experience.

## 1 INTRODUCTION

PROLIFERATION of smart devices, driven by advancements in wireless communication, has led to an exponential growth in data traffic, especially video content, on mobile networks. This surge places significant strain on the backhaul links that connect the base stations to the core network, potentially compromising the user experience [1]. To mitigate this challenge, researchers have explored various strategies to enhance mobile user access to the Internet.

- *M. Rezaei is with the Faculty of Computer Engineering, University of Isfahan, Isfahan, Iran. E-mail: ma.rezaei@eng.ui.ac.ir*
- *A. Montazerolghaem is with the Faculty of Computer Engineering, University of Isfahan, Isfahan, Iran. E-mail: a.montazerolghaem@comp.ui.ac.ir*

Although network densification, bandwidth expansion, and infrastructure upgrades can enhance wireless networks, these solutions may be insufficient in densely populated areas due to the overwhelming traffic demand generated by popular content [1]. To mitigate this problem, caching networks and edge caching have been proposed to reduce content delivery latency and reduce backhaul traffic. Considering the skewed distribution of user requests to a limited amount of popular content, caching this content at different edge locations is considered an effective strategy to reduce network congestion and prevent excessive traffic over backhaul links [2].

Accurate prediction of content popularity in dynamic environments is crucial for effective caching [3]. However, high prediction accuracy does not necessarily guarantee a high cache hit rate, as the latter depends on both demand estimation and network-aware content placement within users' proximity via D2D links. To address this, we employ a supervised machine learning approach based on XGBoost to model user demand. The model leverages features derived from user profiles (e.g., age and occupation), content characteristics (e.g., genre and release year), and historical user–content interactions to predict preference scores, which are subsequently aggregated at the cluster level to estimate content popularity and guide cache placement decisions.

Caching and D2D communication are effective strategies to improve network performance and ensure high-quality service [4]. D2D communication enables direct data exchange between nearby devices without base station involvement, reducing latency, improving throughput, and operating efficiently over short distances with low power consumption [1]. D2D communication also enables efficient content delivery by storing data at the network edge, such as on user devices [5]. Recent works have enhanced D2D caching through advanced optimization, including many-objective content popularity prediction in cloud-edge-end IoT networks [6], joint preference-aware caching and migration in mobile edge networks [7], and energy-efficient fetching via Actor-Critic reinforcement learning [8]. However, they often neglect cluster-level user heterogeneity in prediction.

The excessive use of D2D transmissions can cause harmful interference with the primary mobile network on the same resources. Therefore, in this study, a clustering algo-rithm is used to group nearby D2D users into clusters where multimedia files are exchanged via D2D communications [9]. This clustering also makes it possible to centrally store the preferred content of users in each cluster based on the usage behavior of users in the vicinity of that cluster.

### A. Motivations

In cellular networks, various caching methods have been proposed. This study specifically focuses on caching at user devices, such as smartphones, leveraging the substantial storage capacity of modern devices to store frequently requested multimedia content [10]. However, storage space is limited, necessitating an efficient content selection strategy for caching. Only a small number of requested content is popular and this small number accounts for a large portion of network traffic, while other content is rarely requested. For example, 1% of Facebook videos account for 83% of total watch time [11]. Existing caching methods take advantage of the small amount of popular content and cache them, but content popularity is very dynamic and uncertain. However, it should also be noted that achieving high prediction accuracy does not necessarily ensure high cache hit rates. The hit rate depends not only on predicting user interests but also on how and where the predicted content is placed within the network. Most existing research assumes that all users have the same distribution of content popularity distribution. This model is at best an approximation, as different users actually have different tastes and preferences. Less than 20% of the users generate 80% of multimedia traffic, indicating a heterogeneous user activity level of the users. However, in existing studies, the most active caching approaches ignore user behavior, such as heterogeneous user preferences [3], [12].

### B. Novelty and Contributions

This paper proposes a **Machine Learning-Driven** framework for **Content Popularity Prediction and Cache Optimization (ML-CPCO)** in clustered D2D networks. The main contributions of this work are summarized as follows:

- A *unified ML-driven caching framework* that jointly integrates user clustering, content popularity prediction using *XGBoost*, and *network-aware cache placement*, enabling efficient exploitation of *user heterogeneity* in D2D clustered networks.
- A *cluster-level content popularity modeling* approach that captures heterogeneous user preferences by leveraging *user attributes*, *content features*, and *historical interactions*, while explicitly incorporating *user willingness* to participate in caching.
- A *system-level analysis* demonstrating that improved prediction accuracy alone does not guarantee higher cache hit rates, and showing how the proposed joint design of prediction and placement significantly enhances *cache utilization* and *hit rate* under practical D2D constraints.

Overall, these improvements clarify the novelty of our work and firmly situate it within the broader research on ML-driven caching and D2D optimization.

### C. Organization

The rest of this paper is organized as follows. Section 2 gives a comprehensive survey of the existing literature on content caching in wireless networks. Section 3 presents our proposed content caching framework, including the details of the clustering algorithm and the popularity prediction model. Section 4 evaluates the performance of our proposed system through extensive simulations. Finally, Section 5 concludes the paper and discusses future research directions.

## 2 Related Works

This section provides an overview of the existing research on content caching in wireless networks. These researches can be divided into two categories: Static content popularity-based caching and dynamic content popularity-based caching. The first group assumes that the popularity of content follows a fixed distribution, e.g. Zipf's distribution, and does not propose methods for predicting the popularity of content. The second group recognizes the dynamic nature of content popularity and introduces methods, often based on machine learning, to predict it. These approaches aim to improve cache efficiency and reduce backhaul traffic by adjusting cache placement strategies based on user behavior.

### A. Static Content Popularity Based Caching

The study in [13] proposes a mathematical model to optimize coordinated multipoint transmission using caching in D2D networks, resulting in improved network performance. To address latency and user experience challenges, [14] develops a D2D Collaborative Edge Caching (DCEC) scheme that enhances cache hit probability while reducing delay-associated costs. Reference [15] presents a three-layer hierarchical model for content delivery with correlated content placement that achieves better cache hit rates compared to independent methods. In [16], the authors investigate content placement techniques for D2D caching networks, with experimental results indicating relatively appropriate hit probability performance.

The Research in [17] introduces an incentive-based communication and caching system for D2D clustered networks that improves the cache hit rate and user participation. Reference [1] proposes a D2D cluster-based cache placement algorithm that maximizes the traffic offloaded from the backhaul link, outperforming centralized solutions.

### B. Dynamic Content Popularity Based Caching

In [18], the authors investigate caching strategies in D2D networks using recommendation techniques to improve cache efficiency and reduce latency. Unlike their *recommend to cache* approach, our work adopts a proactive *predict to cache* strategy that directly forecasts user demand without relying on recommendations.

Reference [19] presents an approach to optimize D2D content sharing by using machine learning to predict user behavior and increase the hit rate in D2D caching. The work in [20] formulates caching schemes in D2D networks that consider user preferences and improve network performance. Reference [21] introduces the CoPUP[1] framework in MEC[2], which uses content-based collaborative filtering and CNN[3] to improve the cache hit rate and response time.

1. Content Popularity and User Preferences aware content caching
2. Mobile Edge Computing
3. Convolution Neural Network

Reference [12] proposes a shared caching mechanism that uses LSTM[4] to predict user preferences and cluster users, showing better performance in terms of hit rate, speedup rate, and cache utilization.

Similarly, [6] employs many-objective optimization for content popularity prediction in cache-assisted IoT networks, enhancing hit rates in collaborative edge settings. [7] focuses on joint optimization of preference-aware caching and content migration in cost-efficient mobile edge networks, reducing latency through user-centric strategies. [22] proposes joint optimization of caching and content delivery in air-ground cooperation environments, predicting content popularity dynamically using an LSTM network based on historical data. [8] introduces energy-efficient content fetching in D2D networks using an Actor-Critic reinforcement learning structure, improving resource utilization. Additionally, [23] optimizes joint caching and recommendations from network and user perspectives in wireless D2D networks, boosting overall system efficiency.

[24] presents a deep learning-based caching scheme that improves cache hit probability, backhaul offload, and latency when accessing video content in mobile edge computing networks. In [25], a cooperative caching scheme is proposed with the deployment of MEC clustered servers for the distribution of video content, where the simulation-based evaluation confirms the improvements in streaming service quality. Reference [26] proposes a hybrid caching framework using LSTM for content popularity prediction. Caching is performed at both base stations and selected edge devices to deliver content closer to users. This method improves cache hit rate and reduces latency compared to traditional schemes. In [27], the authors present a novel software-defined approach for joint content popularity forecasting and cache optimization in D2D environments. The proposed system leverages SDN's[5] centralized control and distributed clustering to implement device-level caching strategies that boost hit ratios while minimizing latency.

Most previous research has focused on clustering algorithms and cache placement strategies that rely on predefined content popularity models to maximize backhaul offloading. In contrast, our work presents a novel machine learning-based content popularity prediction method specifically tailored to the dynamic nature of user behavior. Unlike traditional dynamic methods, which may overlook user willingness to participate in caching, our approach accounts for varying user preferences and participation levels, ensuring that only users willing to share content are included in the caching process. Additionally, our clustering mechanism groups users based on proximity and shared content preferences, optimizing the caching process, improving cache hit rates, and minimizing backhaul traffic, thereby providing a more adaptable and efficient solution for content caching.

Traditional caching strategies typically rely on static popularity models (e.g., Zipf) or short-term request statistics, which cannot capture the dynamic and heterogeneous nature of user behavior. Machine learning offers a data-driven alternative capable of modeling temporal variations and complex dependencies between content, users, and con-

4. Long Short-Term Memory
5. Software-Defined Networking

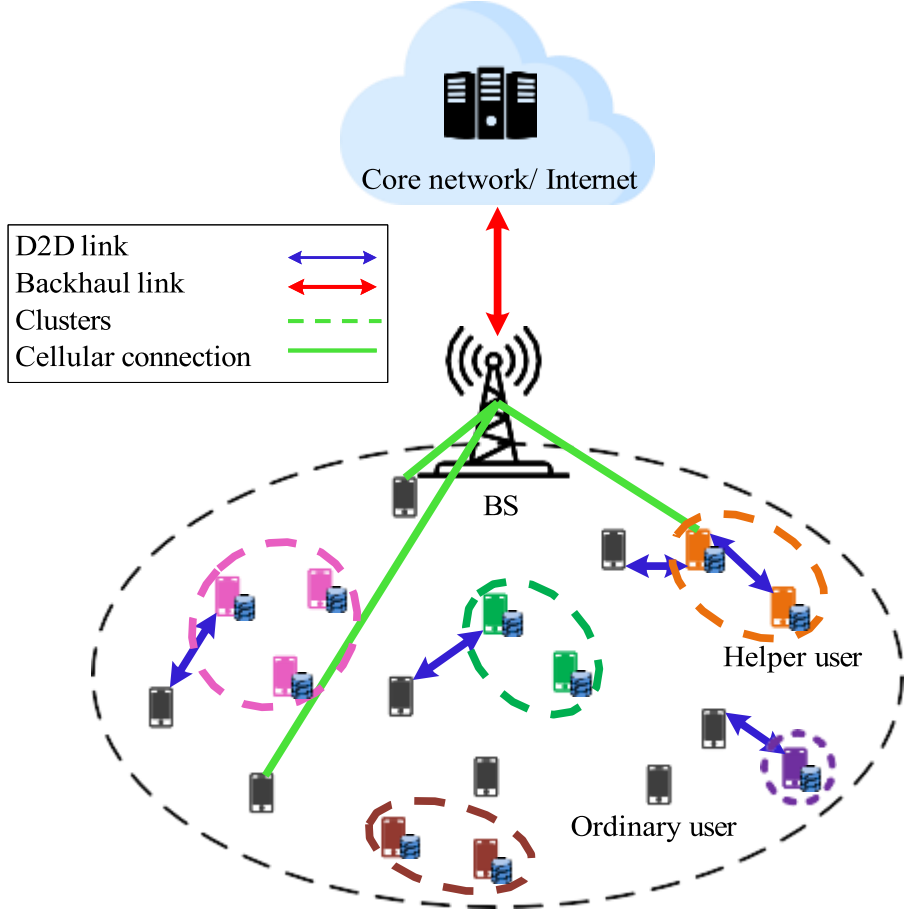

Figure 1. The Proposed ML-CPCO framework

textual factors. By learning from historical request patterns, machine learning-based prediction adapts to changing user interests and diverse content characteristics. In this study, the XGBoost algorithm is employed for its ability to model non-linear relationships and enhance prediction accuracy. Experimental results confirm that our ML-CPCO framework achieves superior cache utilization (approximately 97%) and hit rates compared to baseline approaches, demonstrating its robustness, adaptability, and efficiency in next-generation wireless networks.

## 3 Problem definition

In this study, users share cached multimedia files with nearby devices via D2D communication. Because the distance between the requester and the helper is short, D2D transmission offers efficient content delivery [10]. Each helper stores a limited number of files and directly serves requests for cached content. If the requested file is not found in nearby caches, the request is forwarded to the base station, which retrieves the content through the backhaul and delivers it to the user.

To model this scenario, we consider a cellular network, illustrated in Fig. 1, where a base station serves a set of $U$ users. The users are categorized into two groups: $H$ *helper users*, who are willing to cache and share content via D2D links, and $O$ *ordinary users*, who only request content but do not participate in caching. This user classification, defined in Equation (1), allows us to capture varying levels of user participation and analyze their impact on the overall system performance.

$$U = O + H \tag{1}$$

We used the "MovieLens 1M" dataset [28], which contains 1 million ratings from 6000 users for about 4000 movies. We treat MovieLens user ratings as proxies for content requests, assuming that higher ratings indicate stronger user interest and a higher likelihood of requesting or re-watching the content—an approach commonly used in content popularity prediction and caching studies [21], [29]–[31].

Table 1 summarizes the parameters used in this section. The set $M$ represents the set of contents that can be cached, which in this case is a set of movies. The size of each content ($CS$) is assumed to be one unit, and each user can independently cache up to $CC$ units of content. To make the cache placement problem non-trivial, we assume that $CC < |M|$. The selected dataset consists of three files: the set of users ($U$), the set of movies ($M$), and the set of ratings ($R$). The parameter $MRU$ is created by combining these three sets and is represented as a 2D array containing user information (age and occupation), movie information (genre and year of release) and the rating given by the user to the movie. With D2D communication, two users who are less than $RD$ distance apart can communicate directly, where $RD$ represents the maximum range for D2D communication. If an ordinary user requests content and a nearby helper user is within a distance of $\leq RD$, the helper user can assist if they have the requested content. The parameter $D$ is a 2D array, where $D_{i,j}$ denotes the distance between the users $i$ and $j$ as given by Equation (2).

Table 1
Parameters and Definitions

| Parameter | Definition |
|---|---|
| $U$ | User set |
| $H$ | Helper user set |
| $O$ | Ordinary user set |
| $M$ | Movie set |
| $CS$ | Content size |
| $CC$ | Cache size |
| $R_{u,c}$ | Rating of movie $c$ by user $u$ |
| $MRU$ | Combined information set of users, movies, and rates |
| $RD$ | Maximum D2D communication range |
| $D_{i,j}$ | Distance between user $i$ and user $j$ |
| $A_{i,j}$ | Proximity of user $i$ to user $j$ |
| $Cl$ | Cluster set |
| $PR_{u,c}$ | Predicted rating of movie $c$ by user $u$ |
| $RCl_{cl,c}$ | Predicted rating of movie $c$ by cluster $cl$ |
| $P_{cl,c}$ | Probability of storing content $c$ in cluster $cl$ |
| $X_{u,c}$ | Caching content $c$ at user $u$ |
| $Z_{u,c}$ | Accessing content $c$ by user $u$ from a reachable helper |
| $K^{\max}_{cl,c}$ | Maximum allowed replicas of content $c$ in cluster $cl$ |

$$D_{i,j} = \sqrt{(x_i - x_j)^2 + (y_i - y_j)^2} \tag{2}$$

Additionally, a binary parameter $A$, represented as a 2D array, is defined as the adjacency matrix. $A_{i,j}$ indicates the neighborhood relationship between user $i$ and user $j$, as given by the following Equation:

$$A_{i,j} = \begin{cases} 1 & \text{if } D_{i,j} < RD \\ 0 & \text{otherwise} \end{cases} \tag{3}$$

Helper users are grouped using a clustering algorithm, and the resulting clusters are represented by the parameter $Cl$. To prevent interference with cellular transmissions, D2D links operate in separate frequency bands. Inter-cluster interference is reduced through spatial separation, as the proximity-based clustering algorithm ensures clusters are spaced beyond the D2D range $RD$. Within each cluster, a simple MAC protocol (e.g., CSMA/CA) coordinates transmissions and minimizes intra-cluster interference.

A machine learning model is then used to predict user ratings for movies, with the predictions stored in the 2D array $PR$. The predicted ratings for each movie by the users within each cluster are subsequently calculated and stored in the array $RCl$. Based on the values in $RCl$, a popularity matrix $P$ is defined, where $P_{cl,c}$ represents the probability that content $c$ is cached by the members of cluster $cl$. Assuming that content popularity follows a Zipf distribution, $P_{cl,c}$ is proportional to $\frac{1}{c^{\alpha}}$.

Next, a binary matrix $X$ is generated by executing the cache placement algorithm. If content $c$ is stored in the cache of user $u$, then $X_{u,c} = 1$; otherwise, it equals 0. For ordinary users, all elements in row $u$ of matrix $X$ are set to zero. To capture the accessibility of cached contents, let $Z_{i,c} \in \{0, 1\}$ indicate whether ordinary user $i$ can fetch content $c$ from at least one reachable helper. Each helper user has a limited cache capacity $CC$, and contents may be replicated up to a cluster-specific limit $K^{\max}_{cl,c}$ to preserve content diversity.

As shown in Fig. 1, the base station executes the computational algorithms and manages cache placement. The results are then communicated to the users. The system aims to maximize the number of content items accessed through the caches of the helper users, thereby increasing the hit rate.

The cache placement task can be formulated as a binary optimization problem as follows:

**Objective:**

$$\max_{x,z} \sum_{u\in U} \sum_{c\in M} PR_{u,c}\, Z_{u,c} \tag{4}$$

**Constraints:**

$$Z_{u,c} \leq \sum_{j\in H} A_{u,j} X_{j,c}, \quad \forall u \in U, c \in M \quad \text{(coverage)} \tag{5}$$

$$\sum_{c\in M} X_{j,c} \leq CC, \quad \forall j \in H \quad \text{(cache capacity)} \tag{6}$$

$$X_{j,c} = 0, \quad \forall j \in O, c \in M \quad \text{(helpers only)} \tag{7}$$

$$\sum_{j\in H\cap cl} X_{j,c} \leq K^{\max}_{cl,c}, \quad \forall cl \in Cl, c \in M \quad \text{(replica limit)} \tag{8}$$

$$X_{j,c}, Z_{u,c} \in \{0, 1\}, \quad \forall u \in U, j \in H, c \in M. \tag{9}$$

This formulation explicitly models that all users may generate requests, whereas only helpers cache contents. Constraint (5) guarantees that a user can access a requested file only if at least one reachable helper stores it, Constraint (6) enforces cache size limits, and Constraint (8) restricts content redundancy across clusters. The problem generalizes the Budgeted Maximum Coverage Problem (BMCP), an NP-hard problem; thus, obtaining the exact integer solution requires exponential time in $|H| \cdot |M|$, motivating the use of efficient approximation approaches such as submodular-greedy algorithms.

Building upon this general formulation, the proposed approach exhibits several distinctive characteristics. The optimization objective is driven by *ML-based cluster-level popularity estimates*, enabling adaptation to heterogeneous user demand across clusters. Moreover, the cache placement strategy is explicitly *network-aware*, accounting for practical D2D constraints such as cache capacity, communication range, and helper availability. A greedy solution is adopted to balance performance and computational complexity, providing a scalable and practical approach for large-scale D2D networks.

# 4 PROPOSED METHOD

The proposed methodology, illustrated in Fig. 2, consists of five principal components arranged in a sequential workflow architecture. The processing pipeline begins with dataset ingestion and ends with the generation of cache matrix as system output.

## A. Dataset Preparation

This process involves two stages: First, the $n$ most active users and $k$ top-rated movies are selected from the MovieLens 1M dataset. Since this dataset contains user ratings rather than explicit request logs, each rating is interpreted as a proxy for a potential content request, where higher ratings indicate stronger user interest and higher request likelihood. The output of this stage, consisting of the selected users ($U$), selected movies ($M$), and associated ratings ($R$), is fed into the "Data Preprocessing and Integration" phase. In this phase, data is processed and integrated, combining user attributes (e.g., age, occupation) and movie metadata (e.g., genre, release year). The final result is stored in the variable $MRU$, which serves as the input for the "User Preference Prediction" module. Concurrently, the selected users are also input into the "Clustering Helper Users" module.

## B. Clustering Helper Users

This section also consists of two phases. First, a number of helpers are randomly selected from the previously selected users. Then, in the clustering phase, these helpers are divided into groups using clustering algorithms. The goal of clustering is to facilitate collaboration between helpers and increase the diversity of content within each cluster. In this study, clusters are defined as groups of helpers that have a one-hop distance, allowing them to interact directly with each other and provide services to ordinary users.

Algorithm 1 describes a method for clustering helper users. The algorithm uses a group of helpers ($H$), adjacency information ($A$), and distance metrics ($D$). The algorithm utilizes a nested loop structure. In the outer loop, distinct clusters are formed and each user who joins a cluster is subsequently removed from the set $H$. The algorithm starts by identifying the pair of helper users with the minimum distance. If these users are adjacent, a new cluster is formed. The inner loop then iteratively attempts to integrate as many users as possible into the existing cluster. This process continues until there are no more suitable candidates for inclusion in a cluster. The remaining users who cannot form a cluster are assigned individually to singleton clusters. At the end, the algorithm outputs the resultant set of clusters and the corresponding users within each cluster.

The clustering algorithm naturally limits cluster size and enforces spatial separation among clusters. By grouping only adjacent users within the D2D range $RD$, it reduces the chance of overlapping transmissions and mitigates inter-cluster interference. The bounded cluster size further restricts simultaneous intra-cluster transmissions.

To better understand the algorithm, an example with the following inputs is provided. In this example, the users are distributed over an area of 100 x 100 meters and the $RD$ is set to 80 meters.

$$U = [0, 1, 2, 3, 4, 5] \quad H = [0, 1, 2, 4]$$

**Algorithm 1:** Clustering helper users

```
Input: H, D, A
Output: Cl
1  Initialize loop1 ← 0, loop2 ← 0, g ← −1, Cl ← [];
2  while loop1 = 0 do
3      Find the minimum distance and corresponding
         pair (h1, h2) in H using D;
4      if h1 and h2 are found and they are adjacent then
5          g ← g + 1;
6          Add (h1, h2) to Cl_g;
7          Remove h1 and h2 from H;
8          Initialize loop2 ← 0;
9          while loop2 = 0 do
10             Find set
                 N = {i | i ∈ H, A_{i,h} = 1, ∀h ∈ Cl_g};
11             if N is not empty then
12                 Calculate the sum of distances for each
                     user in N to Cl_g using D;
13                 Find the user in N with the minimum
                     total distance to Cl_g;
14                 Add the found user to Cl_g;
15                 Remove the added user from H;
16             else
17                 loop2 ← 1;
18             end
19         end
20     else
21         loop1 ← 1;
22     end
23 end
24 foreach remaining user in H do
25     Add the user to a new group in Cl;
26 end
27 return Cl;
```

$$D = \begin{bmatrix} 0 & 13 & 131 & 133 & 59 & 140 \\ 13 & 0 & 122 & 124 & 50 & 135 \\ 131 & 122 & 0 & 2 & 73 & 57 \\ 133 & 124 & 2 & 0 & 75 & 56 \\ 59 & 50 & 73 & 75 & 0 & 89 \\ 140 & 135 & 57 & 56 & 89 & 0 \end{bmatrix}$$

$$A = \begin{bmatrix} 1 & 1 & 0 & 0 & 1 & 0 \\ 1 & 1 & 0 & 0 & 1 & 0 \\ 0 & 0 & 1 & 1 & 1 & 1 \\ 0 & 0 & 1 & 1 & 1 & 1 \\ 1 & 1 & 1 & 1 & 1 & 0 \\ 0 & 0 & 1 & 1 & 0 & 1 \end{bmatrix}$$

The results of executing the algorithm for clustering users are presented in Fig. 3, which highlights the clustering of helper users (marked by stars) among ordinary users (gray circles). The algorithm has specifically clustered users 0, 1, 2, and 4.

## C. User Preference Prediction

A two-step process is used: First, the merged $MRU$ dataset is segmented. Then, the training and test datasets are fed into the prediction model. The training data represents historical user requests, while the test data represents future requests. The goal is to predict future user requests, evaluate

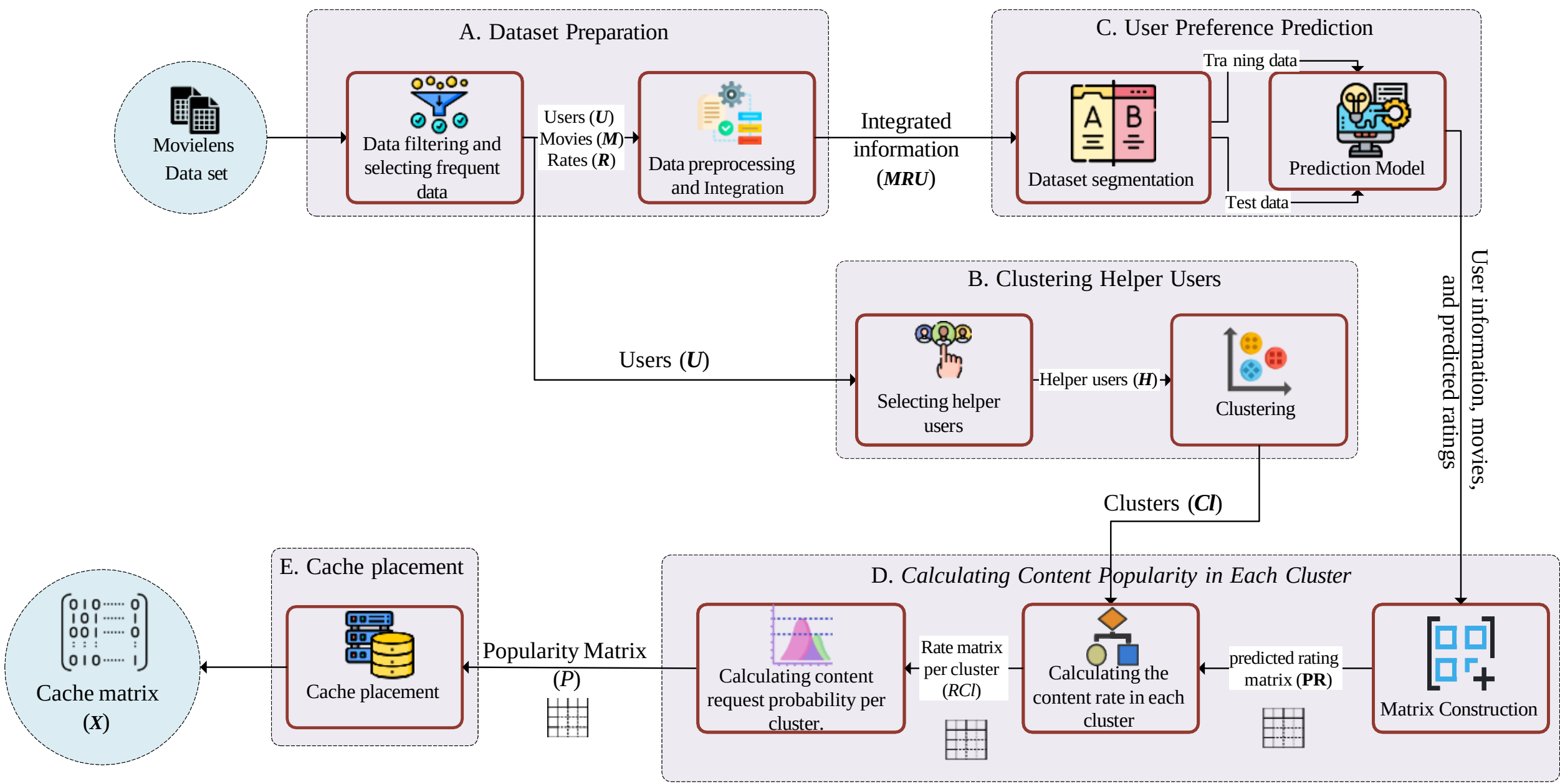

Figure 2. Work flow of the proposed method

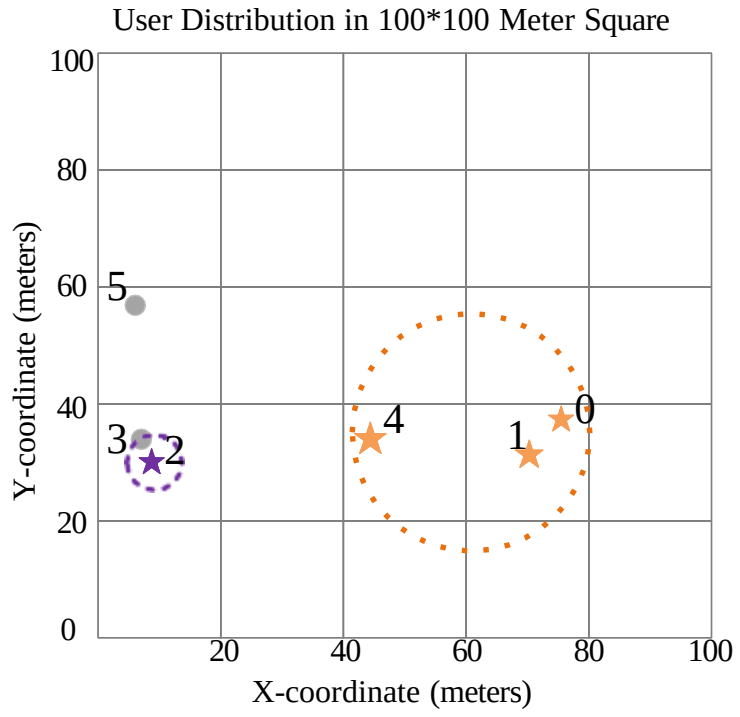

Figure 3. An example of clustering users

the popularity of content within individual clusters, and optimize cache placement.

### D. Calculating Content Popularity in Each Cluster

In this section, after predicting user preferences, the matrix $PR$ is created that contains the predicted ratings for movies. This matrix is then used in Algorithm 2 to calculate the content rating within each cluster. The goal of this algorithm is to assign a user's movie rating directly to the cluster containing the nearest helper user, while distributing a fraction of that rating to neighboring clusters.

Algorithm 2 uses a set of movies ($M$), users ($U$), helpers ($H$), clusters ($Cl$), adjacency matrix ($A$), distance matrix ($D$), and a matrix of predicted ratings for each movie by each user ($PR$) to generate the matrix $RCl$, which represents the ratings of each movie within each cluster. The algorithm operates in two primary loops: for each user and movie, it checks whether the user has provided a rating. If so, the rating is applied to the corresponding clusters associated with the user.

First, the closest neighboring helper to user $u$ is identified and stored in the variable $NH$. If user $u$ is itself a helper, it is stored in $NH$. If there is no neighboring helper for $u$, it's rating does not affect any clusters, and the algorithm proceeds to the next user. If there is a neighboring helper for $u$, the algorithm checks whether $u$'s rating directly or indirectly influences each cluster. For each cluster $cl$, the algorithm considers two cases:

1. If the cluster $cl$ contains the nearest neighboring helper ($NH$): The rating of user $u$ is directly added to the content $c$ in the matrix $RCl$.

2. If $NH$ is not in the cluster: The nearest helper in this cluster is identified, and the rating of user $u$ is applied to $RCl$ relative to the distances.

Finally, the matrix $RCl$ is returned as the output. For example, suppose the inputs are $U = \{0, 1, 2, 3, 4, 5\}$, $H = \{0, 1, 2, 4\}$, $Cl_0 = \{0, 1, 4\}$, and $Cl_1 = \{2\}$ (similar to the previous example), and the parameters $D$ and $A$ have the same values. The $PR$ matrix generated after the "user preference prediction" stage is shown below:

$$PR = \begin{bmatrix} 2 & 0 & 0 & 0 & 4 & 3 & 0 & 0 & 0 & 0 \\ 0 & 0 & 5 & 0 & 0 & 5 & 0 & 0 & 0 & 0 \\ 0 & 0 & 4 & 0 & 0 & 0 & 0 & 5 & 0 & 0 \\ 1 & 0 & 0 & 0 & 0 & 0 & 3 & 0 & 0 & 0 \\ 0 & 0 & 0 & 0 & 0 & 0 & 0 & 0 & 0 & 5 \\ 0 & 0 & 0 & 4 & 0 & 5 & 0 & 0 & 0 & 0 \end{bmatrix}$$

With $M = \{0, 1, 2, 3, 4, 5, 6, 7, 8, 9\}$, the algorithm generates the output matrix $RCl$ as follows:

$$RCl = \begin{matrix} 2.03 & 0.00 & 5.00 & 0.00 & 4.00 & 8.00 & 0.09 & 0.00 & 0.00 & 5.00 \\ 1.00 & 0.00 & 4.00 & 4.00 & 0.00 & 5.00 & 3.00 & 5.00 & 0.00 & 0.00 \end{matrix}$$

Following the calculation of each movie's rating in each cluster and creating the $RCl$ matrix, the popularity of the content within each cluster is determined. This matrix is then fed into the content request probability calculation phase. In this phase, the rating of each movie in each cluster and the sum of the ratings of all movies in each cluster are used to determine the probability of each content request according to Algorithm 3. A higher rating for a

**Algorithm 2:** Calculating content rate in each cluster

```
    Input: M, U, H, Cl, D, A, PR
    Output: RCl
 1  foreach content item c in M do
 2    foreach user u in U do
 3      if PR_{u,c} > 0 then
 4        Find NH, the nearest helper from the set
          H that is adjacent to u;
 5        if NH exists then
 6          foreach cluster cl in Cl do
 7            if NH exists in cl then
 8              Add PR_{u,c} to RCl_{cl,c};
 9            end
10            else
11              Find LocalNH, the nearest
                helper from the set cl that is
                adjacent to u;
12              if LocalNH exists then
13                MinD ← D_{u,NH};
14                LocalD ← D_{u,LocalNH};
15                if LocalD = 0 then
16                  LocalD ← 1;
17                end
18                Calculate Ratio ← MinD / LocalD;
19                Add PR_{u,c} · Ratio to
                  RCl_{cl,c};
20              end
21            end
22          end
23        end
24      end
25    end
26  end
27  return RCl;
```

movie means a higher demand and consequently, greater popularity within the cluster. Algorithm 3, upon receiving the clusters ($Cl$) and the predicted rating matrix for each movie in each cluster ($RCl$), calculates the probability of caching each movie in each cluster. First, the relevant ratings are retrieved from the $RCl$ matrix for each cluster and their sum is calculated. Subsequently, the probability of caching each movie is determined by dividing the rating of each movie by the sum of the ratings and storing the result in the matrix $P$.

The following matrix $P$ is generated after applying Algorithm 3 using the values from the previous examples:

$$P = \begin{bmatrix} 0.08 & 0.00 & 0.21 & 0.00 & 0.17 & 0.33 & 0.00 & 0.00 & 0.00 & 0.21 \\ 0.05 & 0.00 & 0.18 & 0.18 & 0.00 & 0.23 & 0.14 & 0.23 & 0.00 & 0.00 \end{bmatrix}$$

## E. Cache placement

This stage outlines the cache placement process using Algorithm 4. The inputs are the movie set ($M$), the user set ($U$), the cluster set ($Cl$), the adjacency information ($A$), the matrix $P$, and the cache size ($CC$). The output is a binary matrix $X$, indicating whether each content item is stored in a particular user's cache. The core of the algorithm consists of a loop that iterates over each cluster, where

**Algorithm 3:** Calculating the probability of content request in each cluster

```
    Input: Cl, RCl
    Output: P
 1  for g in range (Length of Cl) do
 2    list = RCl[g, :];
 3    sum = sum of all rates in list;
 4    LP = [];
 5    foreach r in list do
 6      LP .append(r/Total);
 7    end
 8    P .append(LP );
 9  end
10  return P ;
```

1. The number of helpers and the total cache capacity are calculated per cluster.

2. The probability of content being cached in a cluster is obtained from the matrix $P$. All zero probabilities are replaced with a small value to account for potential future requests for content that currently appears unlikely. This approach optimizes cache utilization by allowing less likely content to be cached when space is available.

3. Content is ranked in descending order of popularity so that the cache placement process begins with the most requested items.

Lines 7–14 of the algorithm iterate to determine the optimal number of replicas for each content within a cluster. In this step, a proportional replication factor is computed for each content by dividing the popularity values within the set $ClP$. Specifically, the term round($ClP_j/ClP_i$) approximates the relative importance between content items $j$ and $i$, ensuring that contents with higher popularity receive proportionally more replicas across helpers. Since cache storage deals with discrete items, the rounding function converts the ideal proportional ratio into an integer number of replicas while maintaining overall proportionality among contents. This value is then compared with the number of users in the cluster ($N$), and the minimum value is stored in $TC$ to ensure that the number of stored copies does not exceed the number of cluster members. This process continues until the sum of the calculated values is less than the cache capacity. Once the number of replicas for each content is determined, the helpers are sorted based on the number of their neighboring ordinary users. Finally, the content is stored in the cache according to the available storage space and the priority of the helpers.

To illustrate how the algorithm works, a concrete example is presented. The parameters $M$, $U$, $Cl$, $A$, and $P$ are initialized with previously defined values. Additionally, the parameter $CC$ is set to 2. The output of the algorithm, represented by the matrix $X$, is given below:

$$X = \begin{bmatrix} 0 & 0 & 1 & 0 & 0 & 1 & 0 & 0 & 0 & 0 \\ 0 & 0 & 1 & 0 & 0 & 1 & 0 & 0 & 0 & 0 \\ 0 & 0 & 0 & 0 & 0 & 1 & 0 & 1 & 0 & 0 \\ 0 & 0 & 0 & 0 & 0 & 0 & 0 & 0 & 0 & 0 \\ 0 & 0 & 1 & 0 & 0 & 1 & 0 & 0 & 0 & 0 \\ 0 & 0 & 0 & 0 & 0 & 0 & 0 & 0 & 0 & 0 \end{bmatrix}$$

0 1 2 3 4 5 6 7 8 9

**Algorithm 4:** Cache Placement

```
    Input: M, U, Cl, A, P, CC
    Output: X
1   foreach cluster cl in Cl do
2       N = Length of cl;
3       total = N * CC;
4       ClP.append(P_{cl,k}) for k in M;
5       set 10^-6 for values from ClP that are 0;
6       sort ClP in descending order;
7       for i in range (Length of M) do
8           TC = [];
9           TC = [min(N, round(ClP_j / ClP_i))] for j = 0 to
              Length of M-1;
10          sum = Sum of [TC];
11          if sum ≥ total then
12              break;
13          end
14      end
15      SH = Sort helpers in cluster cl according to the
          number of neighboring ordinary users;
16      foreach h in SH do
17          for i in range (Length of TC) do
18              if TC_i > 0 then
19                  sum = Sum of [X_{h,k}] for k in M;
20                  if sum < CC then
21                      m = Get movie corresponding to
                          index i in TC;
22                      X_{h,m} = 1;
23                      Decrement TC_i by 1;
24                  end
25                  else
26                      break;
27                  end
28              end
29          end
30      end
31  end
32  return X;
```

The algorithm concludes with the "cache placement" module. After creating matrix X, user requests are assessed to determine whether they can be fulfilled by the cache. This process starts with a request to the base station, where two key criteria are evaluated (as shown in Fig. 4). If both conditions are satisfied, D2D communication is established, allowing the helper to transfer the content to the requester. Otherwise, the content is retrieved through the backhaul connection.

# 5 EVALUATION

The proposed framework was implemented in Python, using specialized scientific computing libraries. Comprehensive studies were conducted to analyze each component's contribution, followed by comparative evaluations against baseline methods across multiple operational scenarios to validate system efficacy.

## A. Simulation setup

To simulate the proposed approach, a square cellular network consisting of a base station, helper users, and ordinary users is considered. The scenarios were executed using the "MovieLens" dataset, with the main parameters presented in Table 2. The simulation was conducted using Xcode and Python on a system with a 12th Gen Intel Core i7-12650H processor, 16 GB RAM, and a 64-bit operating system.

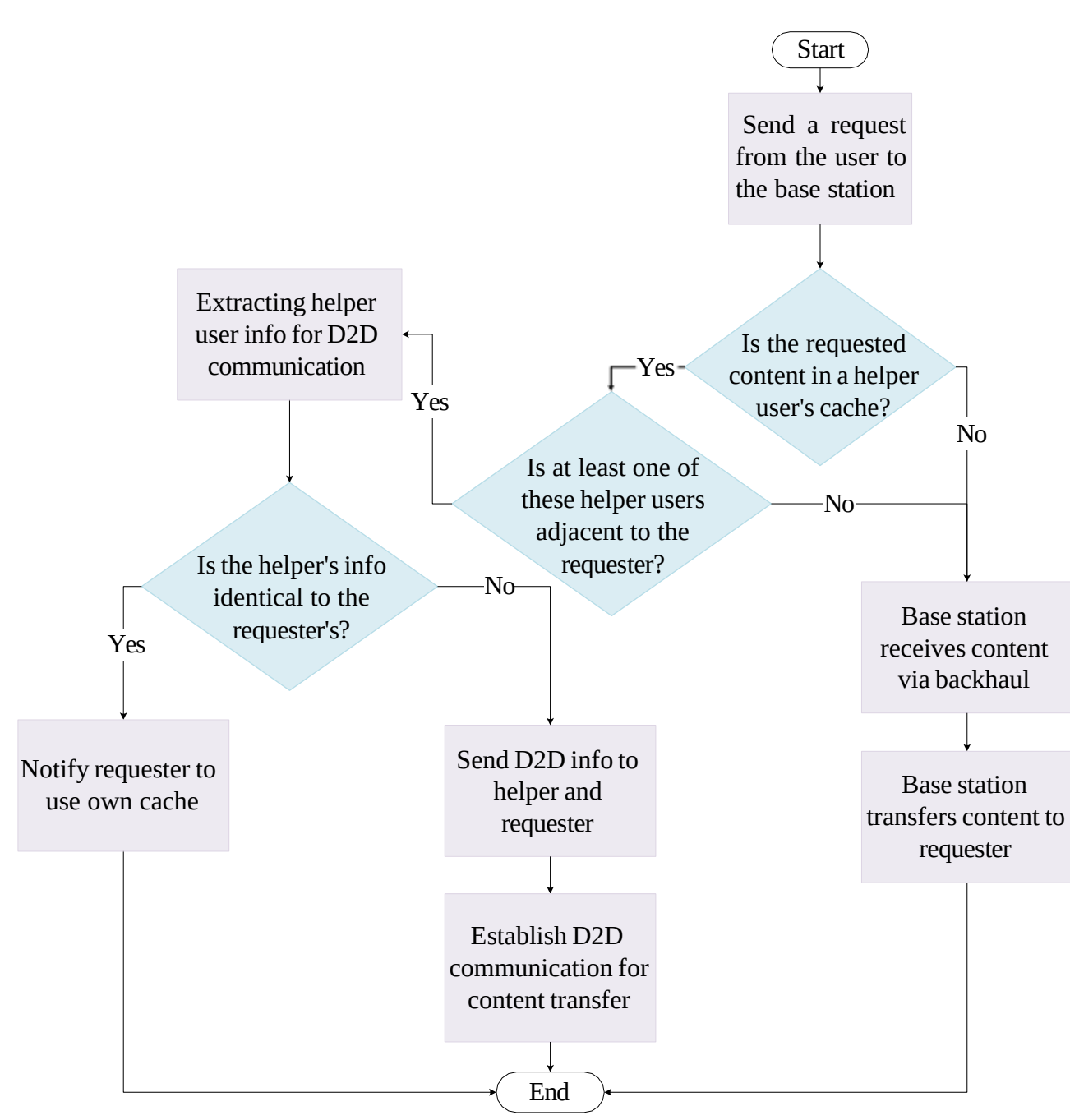

Figure 4. Process of checking user requests after cache placement

Table 2
Simulation Parameters [1]

| Parameter | Value |
|---|---|
| Cell width | 500 m |
| Number of users | 20–220 |
| Number of helper users | 5–100 |
| Maximum D2D communication range | 10–100 m |
| Number of movies (contents) | 600–1100 |
| Cache size per helper | 1–20 units |
| Zipf distribution skewness | 0.1–1.8 |
| Content size | 1 unit |

## B. Choosing a machine learning algorithm

Machine learning algorithms exhibit varying performance across datasets. In this study, the PyCaret library was used to accelerate model development. Table 3 presents the results of applying different machine learning algorithms to the "MovieLens" dataset, indicating XGBoost and LightGBM as the top-performing models. Although XGBoost requires longer execution time, it achieves higher prediction accuracy. While prediction accuracy alone does not directly translate into proportional gains in cache hit rate, XGBoost was selected as the primary model to ensure consistent and reliable evaluations throughout this study.

Crucially, we treat machine learning performance metrics and system-level metrics (i.e., cache hit rate) as distinct optimization objectives. While machine learning accuracy quantifies how precisely the model reproduces users' historical preferences, the cache hit rate depends on (i) predicted demands, (ii) the spatial distribution of helpers and their caches, and (iii) the underlying placement policy.

From an analytical perspective, cache hit rate depends mainly on preserving the relative popularity ordering of the most frequently requested contents within each cluster, rather than on overall prediction accuracy across all items. Due to limited cache capacity, accurate predictions for low-demand contents do not contribute to cache hits, whereas correctly identifying top popular contents—even with moderate accuracy—can yield higher hit rates. By aggregating user-level predictions into cluster-level popularity and optimizing cache placement accordingly, the proposed framework directly aligns learning outputs with the objective of maximizing cache hit rate.

Table 4 shows the results of the proposed method for a cellular network with 100 active users and 700 frequently rated movies. In this experiment, 60 users are employed as helper users, and the maximum D2D communication range is set to 50 m, with cache capacity varying from 1 to 10 units. Different machine learning models are used to predict user preferences.

The perfect accuracy (100%) in the last row of Table 4 indicates that real data are used for cache placement in that row. This scenario serves as a benchmark for evaluating the performance of the proposed method. The results in Tables 3 and 4 show slight differences, as Table 3 considers the entire dataset, whereas Table 4 focuses on only 100 of the most active users and 700 top-rated movies. Nevertheless, XGBoost is the most accurate model. As cache capacity increases, the hit rate also improves. This increase is observed in all machine learning models. By storing more content in the cache, the probability of fulfilling requests from the cache increases, leading to an overall improvement in system performance.

In Table 4, the highest hit rate corresponds to the real data (test dataset), and among the specified models, the one with the highest hit rate is highlighted in red. A comparison between model accuracy and hit rates indicates that caching performance is not directly dependent on prediction accuracy; even models with moderate accuracy (≈ 40%) can achieve high hit rates. This is because the modular design of the proposed ML-CPCO framework—particularly its cluster-level popularity estimation and network-aware cache placement—effectively compensates for minor prediction errors. Therefore, XGBoost was selected for further evaluation as it provides the highest average hit rate and stable overall performance.

## C. Performance Evaluation Metrics

In this research, performance evaluation focuses on three main metrics: cache hit rate, cache utilization, and execution time.

The Cache Hit Rate ($CHR$) [21] is calculated using the following formula:

$$CHR = \frac{CH}{CH + CM} \tag{10}$$

where:

- $CH$ is the number of cache hits,
- $CM$ is the number of cache misses.

Cache utilization indicates the proportion of the cache capacity being utilized [12]. The cache utilization of helper user $h$ is calculated by dividing the number of requests served from the cache by the capacity of the helper's cache, as shown in Equation (5).

$$CU_h = \frac{Hit_Count}{Cache_Capacity} \tag{11}$$

Overall cache utilization is calculated by averaging across all helper users, as shown in Equation (6).

$$CU = \frac{\sum_{h \in H} CU_h}{|H|} \tag{12}$$

Runtime measures algorithm performance by indicating task execution time, which helps assess efficiency and optimize algorithms.

## D. Evaluation of content rate calculation methods in each cluster

This study investigates the calculation of the content rate within each cluster as a key component of the proposed methodology. To evaluate the effectiveness of the proposed approach, it is compared with three alternative methods:

- Approach 1: Only helper ratings are applied with a weight of 1 to their own cluster; ordinary user ratings do not affect any cluster.
- Approach 2: A user's rating affects only the cluster containing the nearest helper. If the rater is a helper, their rating is applied (with a weight of 1) directly to their own cluster. Otherwise, the nearest helper to the rater is identified, and the given rating is applied (with a weight of 1) to that helper's cluster.
- Approach 3: Each user's rating is applied with a weight of 1 to all clusters that have at least one helper neighbor to the rater.
- Approach 4 (proposed): A user's rating is applied with a weight of 1 to the cluster of the nearest neighboring helper, and is also applied to other clusters with neighboring helpers using a coefficient proportional to their proximity.

To assess the performance of each method, two experimental scenarios are considered, with the details presented in Table 5. In scenario 1, 100 active users and their top 700 rated movies were selected. 40 users were randomly chosen, with a maximum D2D communication range of 30 meters. Fig. 5 shows the user distribution in a 500 x 500 meter area, with helper users grouped into 5 clusters.

Fig. 6 shows cache hit rate results for four approaches to calculating content rates in each cluster. As cache capacity increases from 1 to 10 units, the hit rate improves for all approaches because a larger cache increases the likelihood of storing diverse content. The fourth approach achieves the highest performance, while the first approach performs the weakest.

In the first approach, ratings are applied only to helper users, limiting the number of users with caching ability and resulting in poor hit rate performance. The third approach applies each user's rating to all clusters with at least one neighboring helper, but performs worse than the second approach due to redundant storage and reduced content diversity. Since only one neighboring helper's cache is needed to fulfill a request, other caches remain underutilized.

The fourth approach significantly outperforms the others by considering the rates of all users in each cluster's rate calculation. Unlike the second approach, which affects only one cluster, the fourth approach applies the rating to all clusters with a neighboring helper user. It also avoids the inefficiencies of the third approach by applying rates relative to the user's proximity to each cluster, preventing redundant content caching.

Table 3
The Result Of Running Machine Learning Algorithm On "Movielens" Dataset.

| Model | Accuracy | AUC | Recall | Prec. | F1 | Kappa | MCC | TT (Sec) |
|---|---|---|---|---|---|---|---|---|
| Extreme Gradient Boosting | 0.3903 | 0.6444 | 0.3903 | 0.3969 | 0.3382 | 0.1106 | 0.1271 | 18.6440 |
| Light Gradient Boosting Machine | 0.3799 | 0.6306 | 0.3799 | 0.3967 | 0.3051 | 0.0807 | 0.1051 | 47.2130 |
| Gradient Boosting Classifier | 0.3603 | 0.5927 | 0.3603 | 0.3847 | 0.2482 | 0.0341 | 0.0586 | 241.0500 |
| Ada Boost Classifier | 0.3520 | 0.5651 | 0.3520 | 0.2923 | 0.2307 | 0.0179 | 0.0326 | 11.8690 |
| Ridge Classifier | 0.3505 | 0.0000 | 0.3505 | 0.2813 | 0.2201 | 0.0125 | 0.0256 | 0.4270 |
| Linear Discriminant Analysis | 0.3504 | 0.5669 | 0.3504 | 0.2878 | 0.2334 | 0.0184 | 0.0324 | 1.9190 |
| Logistic Regression | 0.3499 | 0.5056 | 0.3499 | 0.1224 | 0.1814 | 0.0000 | 0.0000 | 2.1160 |
| Dummy Classifier | 0.3499 | 0.5000 | 0.3499 | 0.1224 | 0.1814 | 0.0000 | 0.0000 | 0.2860 |
| Random Forest Classifier | 0.3498 | 0.6106 | 0.3498 | 0.3470 | 0.3482 | 0.1221 | 0.1221 | 74.8150 |
| K Neighbors Classifier | 0.3475 | 0.6003 | 0.3475 | 0.3482 | 0.3449 | 0.1169 | 0.1174 | 95.4840 |
| Naive Bayes | 0.3450 | 0.5079 | 0.3450 | 0.2114 | 0.1974 | 0.0011 | 0.0031 | 0.5420 |
| Decision Tree Classifier | 0.3444 | 0.5656 | 0.3444 | 0.3477 | 0.3458 | 0.1231 | 0.1232 | 2.8290 |
| Extra Trees Classifier | 0.3430 | 0.5856 | 0.3430 | 0.3438 | 0.3433 | 0.1179 | 0.1179 | 66.8780 |
| Quadratic Discriminant Analysis | 0.2929 | 0.5595 | 0.2929 | 0.3055 | 0.2927 | 0.0612 | 0.0620 | 0.7750 |
| SVM - Linear Kernel | 0.2406 | 0.0000 | 0.2406 | 0.0639 | 0.0999 | 0.0000 | 0.0000 | 296.1720 |

Table 4
Hit Rate In Different Machine Learning Models With Different Cache Size

| Model / Cache Capacity | 1 | 2 | 3 | 4 | 5 | 6 | 7 | 8 | 9 | 10 | Accuracy | Average hit rate |
|---|---|---|---|---|---|---|---|---|---|---|---|---|
| Extreme Gradient Boosting | 0.075889 | 0.135435 | 0.193406 | 0.245364 | 0.292566 | 0.338467 | 0.382894 | 0.420844 | 0.460912 | 0.499192 | 0.455629 | 0.304497 |
| Light Gradient Boosting Machine | 0.07541 | 0.136956 | 0.193019 | 0.244661 | 0.292829 | 0.337381 | 0.382478 | 0.420858 | 0.460041 | 0.498278 | 0.452471 | 0.304191 |
| Gradient Boosting Classifier | 0.072884 | 0.136181 | 0.190952 | 0.241273 | 0.289614 | 0.336835 | 0.378976 | 0.418992 | 0.459037 | 0.500689 | 0.432399 | 0.302543 |
| Ada Boost Classifier | 0.07277 | 0.133971 | 0.189832 | 0.240757 | 0.285595 | 0.334281 | 0.377253 | 0.417269 | 0.458663 | 0.497646 | 0.386857 | 0.300804 |
| Ridge Classifier | 0.073946 | 0.133655 | 0.189029 | 0.239293 | 0.286916 | 0.334711 | 0.378172 | 0.421059 | 0.462137 | 0.49779 | 0.377687 | 0.301671 |
| Linear Discriminant Analysis | 0.073344 | 0.134172 | 0.188311 | 0.240642 | 0.287031 | 0.333563 | 0.375761 | 0.418619 | 0.460214 | 0.498163 | 0.373714 | 0.300982 |
| Logistic Regression | 0.072511 | 0.134889 | 0.187421 | 0.23757 | 0.288236 | 0.332329 | 0.378746 | 0.418848 | 0.458864 | 0.499053 | 0.349159 | 0.300847 |
| Dummy Classifier | 0.072511 | 0.134889 | 0.187421 | 0.23757 | 0.288236 | 0.332329 | 0.378746 | 0.418848 | 0.458864 | 0.499053 | 0.349159 | 0.300847 |
| Random Forest Classifier | 0.073372 | 0.136583 | 0.18989 | 0.240929 | 0.288351 | 0.333333 | 0.375416 | 0.416236 | 0.455506 | 0.494087 | 0.407132 | 0.30037 |
| K Neighbors Classifier | 0.074951 | 0.136698 | 0.193334 | 0.245551 | 0.292456 | 0.336922 | 0.380067 | 0.420398 | 0.460075 | 0.498163 | 0.394193 | 0.303862 |
| Naive Bayes | 0.073401 | 0.134717 | 0.187105 | 0.238546 | 0.288782 | 0.332013 | 0.377512 | 0.419078 | 0.458951 | 0.497818 | 0.350178 | 0.300792 |
| Decision Tree Classifier | 0.072454 | 0.135549 | 0.192674 | 0.245264 | 0.293719 | 0.33784 | 0.377741 | 0.418705 | 0.45941 | 0.497617 | 0.348854 | 0.303097 |
| Extra Trees Classifier | 0.073774 | 0.132621 | 0.187852 | 0.242393 | 0.289987 | 0.33451 | 0.373464 | 0.414772 | 0.454214 | 0.494345 | 0.3784 | 0.299793 |
| Quadratic Discriminant Analysis | 0.066282 | 0.121627 | 0.177001 | 0.224854 | 0.270582 | 0.316225 | 0.358709 | 0.397807 | 0.438942 | 0.475772 | 0.314926 | 0.28478 |
| Real | 0.077219 | 0.138592 | 0.19609 | 0.249168 | 0.298829 | 0.34539 | 0.391865 | 0.430934 | 0.470605 | 0.508124 | 1 | 0.310681 |

Table 5
Scenario Parameters Table

| Scenario | Users ($U$) | Helper Users ($H$) | Videos ($M$) | D2D Range ($RD$) | Cache Size ($CC$) |
|---|---|---|---|---|---|
| 1 | 100 | 40 | 700 | 30 m | {2,4,6,8,10} |
| 2 | 100 | {10,20,30,40,50,60,70,80,90,100} | 700 | 30 m | 4 |

In Scenario 2, increasing the number of helper users (from 10 to 100) improves the cache hit rate, as shown in Fig. 7. A larger number of helper users expands the overall content storage capacity and allows for a more efficient distribution of cached content, thereby enhancing access to requested content. As shown in Fig. 7, the first approach performs poorly with fewer helper users but improves as their number increases, since it considers only helper ratings within clusters. When there are 100 helper users and no ordinary users, its performance matches that of the second approach. The fourth approach also performs similarly to the first and second approaches when all users are helpers.

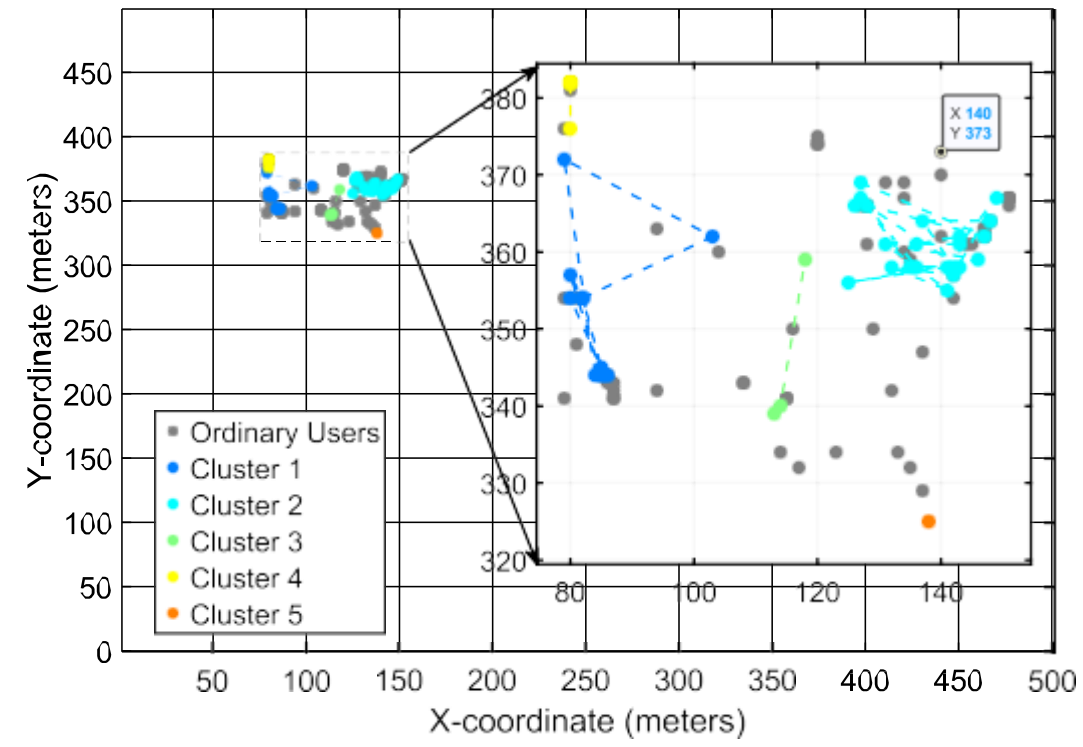

Figure 5. Clustering 100 active and 40 helper users in 500×500

Fig. 8 shows the case where all 100 users are helpers, with each belonging to a separate cluster. Under these conditions, in the first and second approaches, the rate of each user is applied only within their respective cluster. The fourth approach behaves similarly but slightly adjusts the rates in surrounding clusters by applying relative weights. However, when the scoring user is a helper, the impact on neighboring clusters remains negligible and can be ignored.

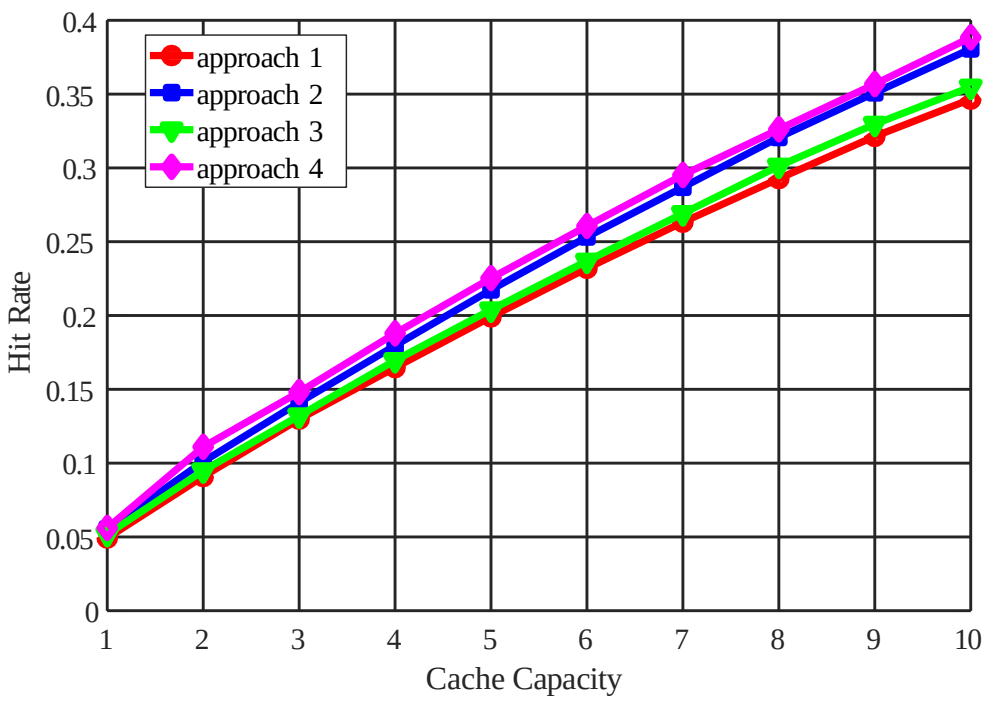

Figure 6. Comparing cache hit rates under cache resizing for different approaches to calculating content rating per cluster

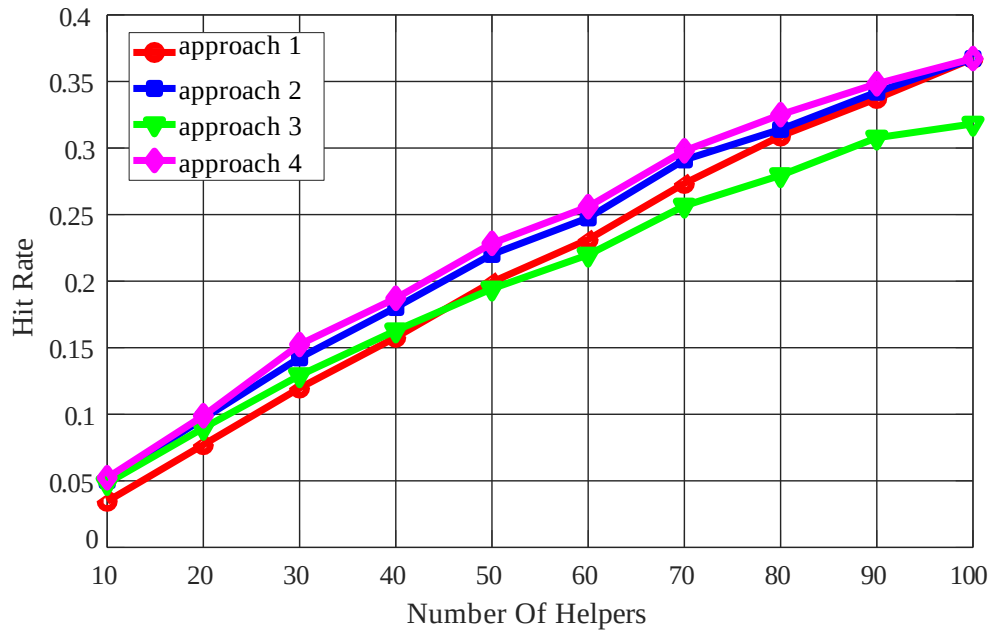

Figure 7. Comparison of cache hit rates under varying number of helper users for different approaches to calculating content rating per cluster

Based on the above results, the fourth approach demonstrates the best performance among all methods. Therefore, it is selected to calculate content rates in each cluster in this study. The remainder of this section presents results using the fourth approach for content rate calculation in each cluster.

## E. Comparison of the proposed approach with previous approaches

The proposed method is evaluated by comparing it with the approach in [1] and common cache placement policies such as LFU[6], FIFO[7], and popularity-based methods. In LFU, the contents with the least usage frequency are removed from the cache. FIFO removes the earliest content added to the cache, while the popularity-based approach stores the most-viewed content. These methods are compared based on cache hit rate, cache efficiency, and execution time in seven scenarios with varying parameters, such as the number of users, the number of movies, cache size, and D2D communication range. The parameters for these scenarios are presented in Table 6.

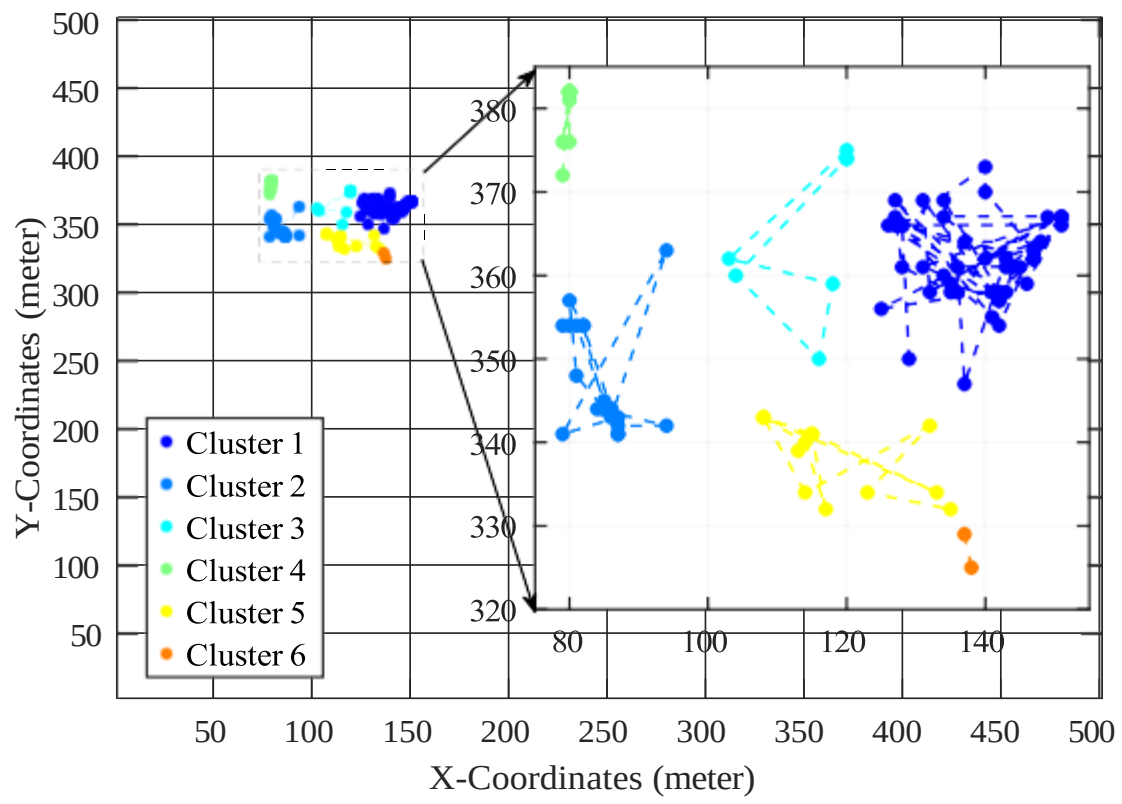

Figure 8. Clustering 100 active and helper users in 500×500

The Scenario 1 results in Fig. 9 show that the proposed method consistently attains the highest hit rate, with the advantage becoming more pronounced as cache capacity increases. The popularity-based method yields the lowest performance due to its simple caching strategy that ignores user preferences, while LFU and FIFO also achieve lower hit rates for similar reasons. The proposed method additionally offers superior cache efficiency, and comparing Fig. 9(a) and Fig. 9(b) confirms that higher hit rates correspond to more efficient cache utilization.

In terms of execution time, the proposed method performs comparably to the approach in [1]. Although the popularity-based method has a shorter execution time, it results in lower hit rates and cache efficiency. LFU and FIFO, however, exhibit higher execution times.

Fig. 10 shows that increasing the number of helper users from 30 to 60 significantly improves the average hit rate in the proposed method (60.97%) and in method [1] (62.14%), compared to smaller increases in LFU (16.43%), FIFO (28.85%), and popularity-based (8.63%) methods. With 60 helpers and a cache size of 20 units, the proposed method achieves a hit rate 80%, reducing backhaul traffic.

In Scenario 2, the impact of changing the skewness coefficient on the performance of the proposed method is investigated. By varying the skewness coefficient of the Zipf distribution from 0.1 to 1.8, the evaluation results are presented in Fig. 11. According to Fig. 11(a), as the skewness coefficient changes, only the hit rate of the research approach in [1] varies, while the other methods maintain consistent performance. As the skewness coefficient increases, the hit rate in method [1] decreases, reaching a level similar to the LFU method at a skewness coefficient of 1.8. However, the proposed method continues to outperform. Fig. 11(b) also shows that cache efficiency changes only in method [1], decreasing as the skewness coefficient increases. In Fig. 11(c), the execution time of method [1] increases with increasing skewness.

In Scenario 3, the skewness coefficient is set to 0.1, which yields the best hit rate and cache efficiency for the approach in [1]. The performance of all methods is evaluated for different numbers of movies, ranging from 600 to 1100, as shown in Fig. 12. In Fig. 12(a), increasing the number of movies reduces the hit rate in both the proposed method and the approach in [1], since greater request diversity lowers the chance of serving content from helper caches. The other three methods do not exhibit this decline because they do not incorporate user preferences, though the proposed method still achieves the highest hit rate. Fig. 12(b) shows

6. Least Frequently Used
7. First In First Out

Table 6
Parameters values in different scenarios

| Parameter | Scenario 1 | Scenario 2 | Scenario 3 | Scenario 4 | Scenario 5 | Scenario 6 | Scenario 7 |
|---|---|---|---|---|---|---|---|
| Users ($U$) | 100 | 100 | 100 | 100 | {20, 40, ..., 220} | {20, 40, ..., 200} | 100 |
| Ordinary Users ($O$) | 70 | 70 | 70 | {90, 80, ..., 0} | {15, 30, ..., 165} | {10, 20, ..., 100} | 70 |
| Helper Users ($H$) | 30 | 30 | 30 | {10, 20, ..., 100} | {5, 10, ..., 55} | {10, 20, ..., 100} | 30 |
| Movies ($M$) | 700 | 700 | {600, 650, ..., 1100} | 700 | 700 | 700 | 700 |
| Maximum D2D Range ($RD$) | 30 m | 30 m | 30 m | 30 m | 30 m | 30 m | {10, 20, ..., 100} |
| Cache Capacity ($CC$) | {2, 4, ..., 20} | 6 | 6 | 6 | 6 | 6 | 6 |
| Zipf Skewness | 1 | {0.1, ..., 1.8} | 0.1 | 0.1 | 1/0 | 0.1 | 0.1 |

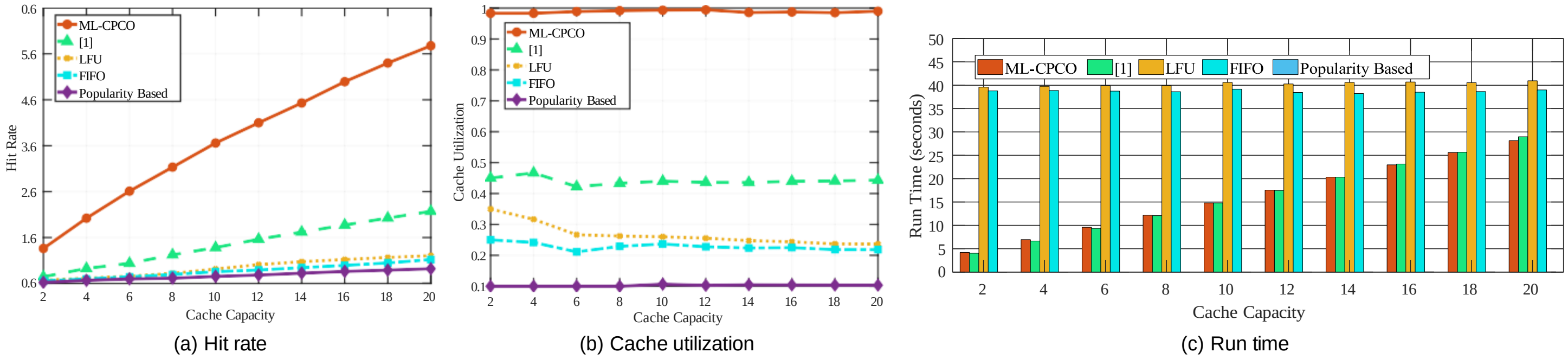

(a) Hit rate (b) Cache utilization (c) Run time

Figure 9. Comparison of different evaluation criteria under cache size change in scenario 1

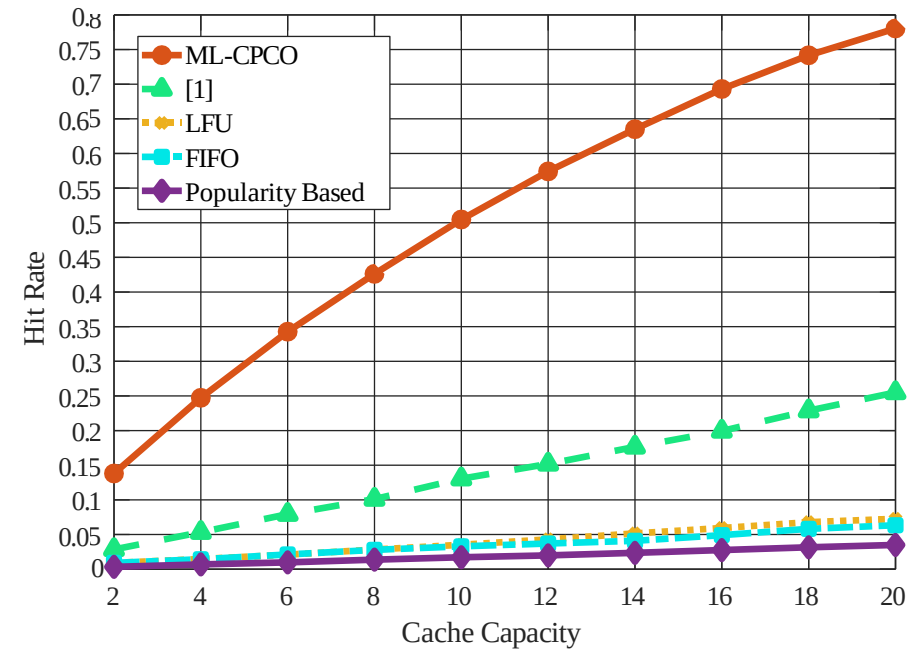

Figure 10. Comparison of hit rates under cache size change in scenario 1 with 60 helper users

that increasing the number of movies has little impact on cache efficiency, with the proposed method maintaining the highest efficiency. In Fig. 12(c), the execution time of all methods increases as the number of movies grows.

Scenario 4 analyzes the effect of increasing the number of helper users (from 10 to 100) while reducing the number of ordinary users (from 90 to 0). Fig. 13(a) shows that as the number of helper users increases, the hit rate significantly improves in both the proposed method and the method in [1], reaching about 50% with 100 helper users. The other methods (LFU, FIFO, and popularity-based) show no improvement. This is because the proposed method and the method in [1] leverage clustering and user behavior to make better use of the increased number of helper users and cache memory, while the other methods, which are based on fixed criteria, cannot fully benefit from these changes.

In Fig. 13(b), the proposed method offers the highest efficiency for all values of helper users compared to the other methods. The approach in [1] also achieves higher efficiency than LFU, FIFO, and the popularity-based method. However, in these three methods, as the number of helper users increases, cache efficiency decreases because their low hit rate causes the cached content to be used less and allocated to less frequently used content. Fig. 13(c) shows the execution time of different methods as the number of helper users increases. An increase in the number of helper users results in a larger cache size and a greater number of processing operations, ultimately extending the execution time of all methods.

In Scenario 4, the methods are evaluated with varying numbers of helper users, while in Scenario 5, the total user count (ranging from 20 to 220) is varied, maintaining a constant ratio of 25% helper users and 75% normal users. Fig. 14(a) shows that as the number of users and helper users increases, the hit rate of the proposed method improves due to two factors: enhanced data accuracy from a larger user base and increased cache availability with more helpers. Although the variety of requests may increase with a growing number of users, a limited number of popular files still account for most requests, thereby boosting the hit rate. This trend is also observed in the approach presented in [1]. In contrast, methods such as LFU, FIFO, and popularity-based approaches show minimal impact on the hit rate when the number of users and helper users changes. This is because these methods do not base content caching on actual user request patterns but instead rely on fixed, predetermined assumptions. Consequently, variations in the number of users and their request patterns do not significantly affect the hit rate of these methods.

Fig. 14(b) shows that the proposed method achieves the highest cache efficiency despite variations in the number of users, making optimal use of the cache. In comparison, LFU, FIFO, and popularity-based methods demonstrate lower efficiency. Fig. 14(c) displays the execution time of the methods as the number of users increases. As the number of users increases, along with a corresponding rise in helper users, the available cache size expands. This leads to a higher number of processing operations in each of the examined methods. Consequently, the execution time for all methods also increases.

In Scenario 6, the helper-user ratio is raised to 50%, and the results are shown in Fig. 15. As seen in Fig. 15(a), all methods follow the same trend as before, but the hit rate increases at all stages due to the larger number of helper users. In Fig. 15(b), the proposed method maintains the highest cache efficiency, while the popularity-based method remains the lowest. Fig. 15(c) shows higher execution times for all methods because more helper users increase cache size and processing, though this increase is smaller for the proposed method and the approach in [1] compared to LFU and FIFO.

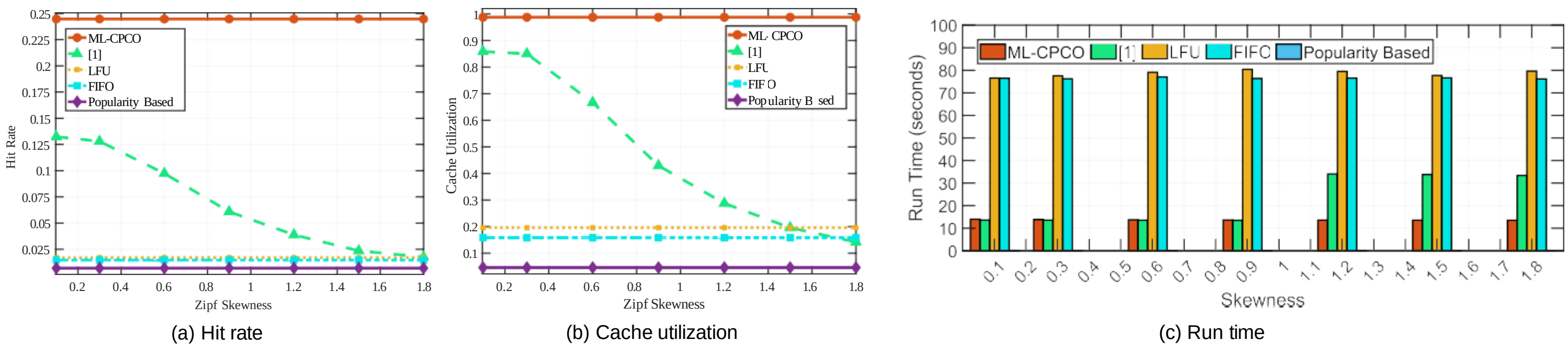

(a) Hit rate (b) Cache utilization (c) Run time

Figure 11. Comparison of different evaluation criteria under changing skewness coefficient in scenario 2

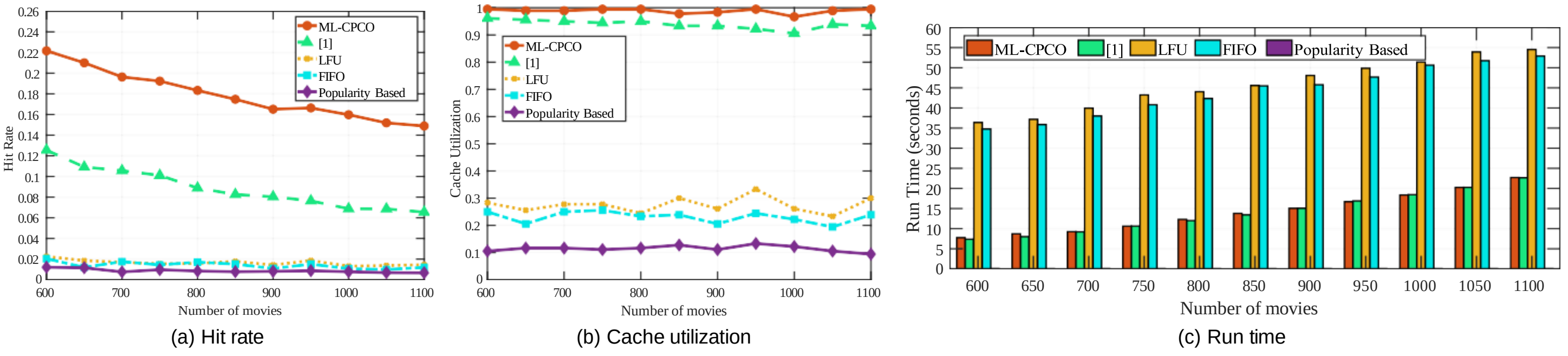

(a) Hit rate (b) Cache utilization (c) Run time

Figure 12. Comparison of different evaluation criteria under changing number of films in scenario 3

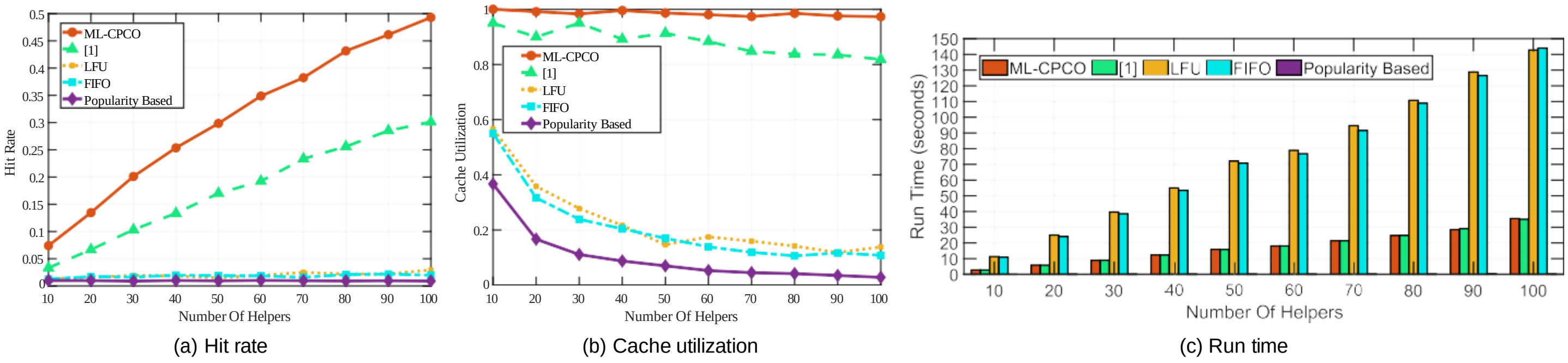

(a) Hit rate (b) Cache utilization (c) Run time

Figure 13. Comparison of different evaluation criteria under changing number of helper users in scenario 4

In Scenario 7, the maximum D2D communication range ($RD$) varies from 10 to 100 meters to evaluate all methods. Figure 16 illustrates the user distribution within a 500 m $\times$ 500 m cell and the resulting clusters for different $RD$ values. At $RD$ = 10 m, the number of clusters is maximal and each cluster is small. As $RD$ increases, clusters become fewer and larger. At $RD$ = 90 m, all helper users form a single cluster, and increasing $RD$ to 100 m causes no further changes, as no additional cluster merges are possible.

Fig. 17 presents the evaluation results for Scenario 7. In Fig. 17(a), it can be observed that in both the proposed method and the approach presented in [1], which utilize clustering for cache placement, the hit rate increases as $RD$ increases. However, in the LFU, FIFO, and popularity-based methods, the hit rate does not change significantly. The increase in the hit rate observed in the proposed and [1] methods can be attributed to two factors:

1. Increased content diversity: Clustering enhances content diversity by grouping similar items together, allowing a broader range of relevant content to be discovered within each cluster.

2. Increased helper users: As $RD$ increases, the number of helper users who can provide content to the requesting user also increases.

In Fig. 17(b), it can be seen that the proposed method has the highest efficiency, while LFU, FIFO, and popularity-based methods show a decreasing trend in efficiency with increasing $RD$. This behavior is due to the lack of content placement based on clustering in these methods, such that similar content is stored in the cache of all helper users. Therefore, with an increase in $RD$, the number of helper users whose cache is used decreases, and this reduces cache efficiency. Fig. 17(c) shows that with increasing $RD$, the execution time of the methods does not change significantly.

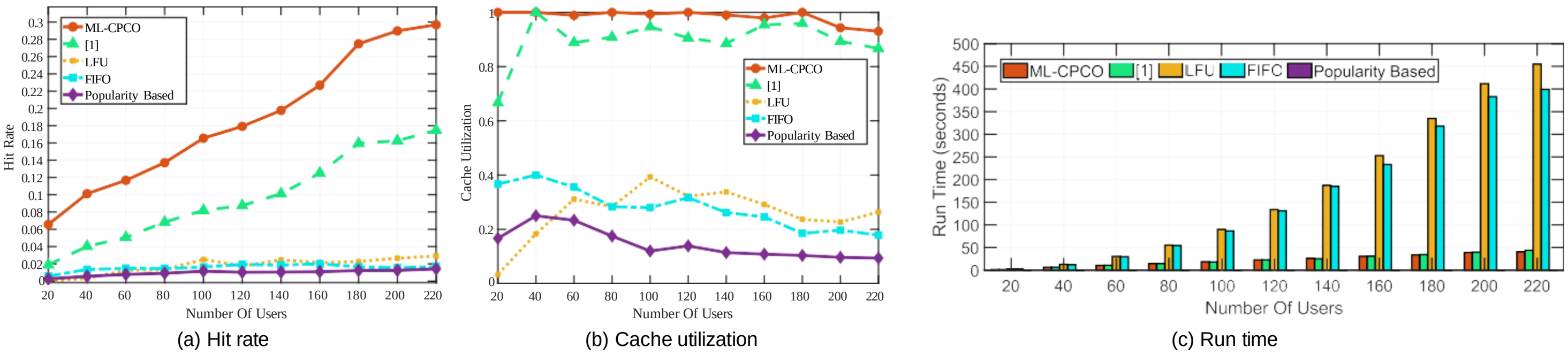

Figure 14. Comparison of different evaluation criteria under changing the number of ordinary users and the number of helper users with a ratio of 25% in scenario 5

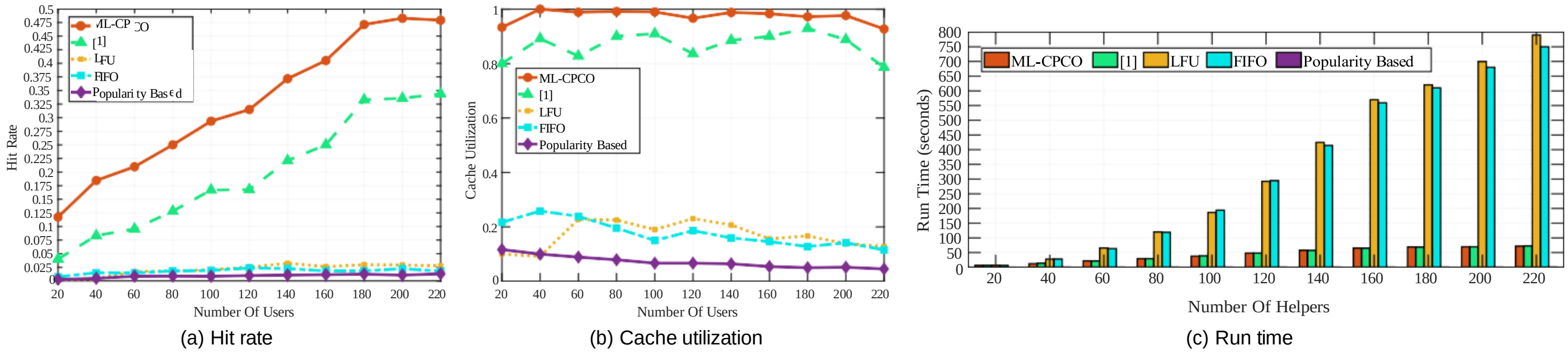

Figure 15. Comparison of different evaluation criteria under changing the number of ordinary users and the number of helper users with a ratio of 50% in scenario 6

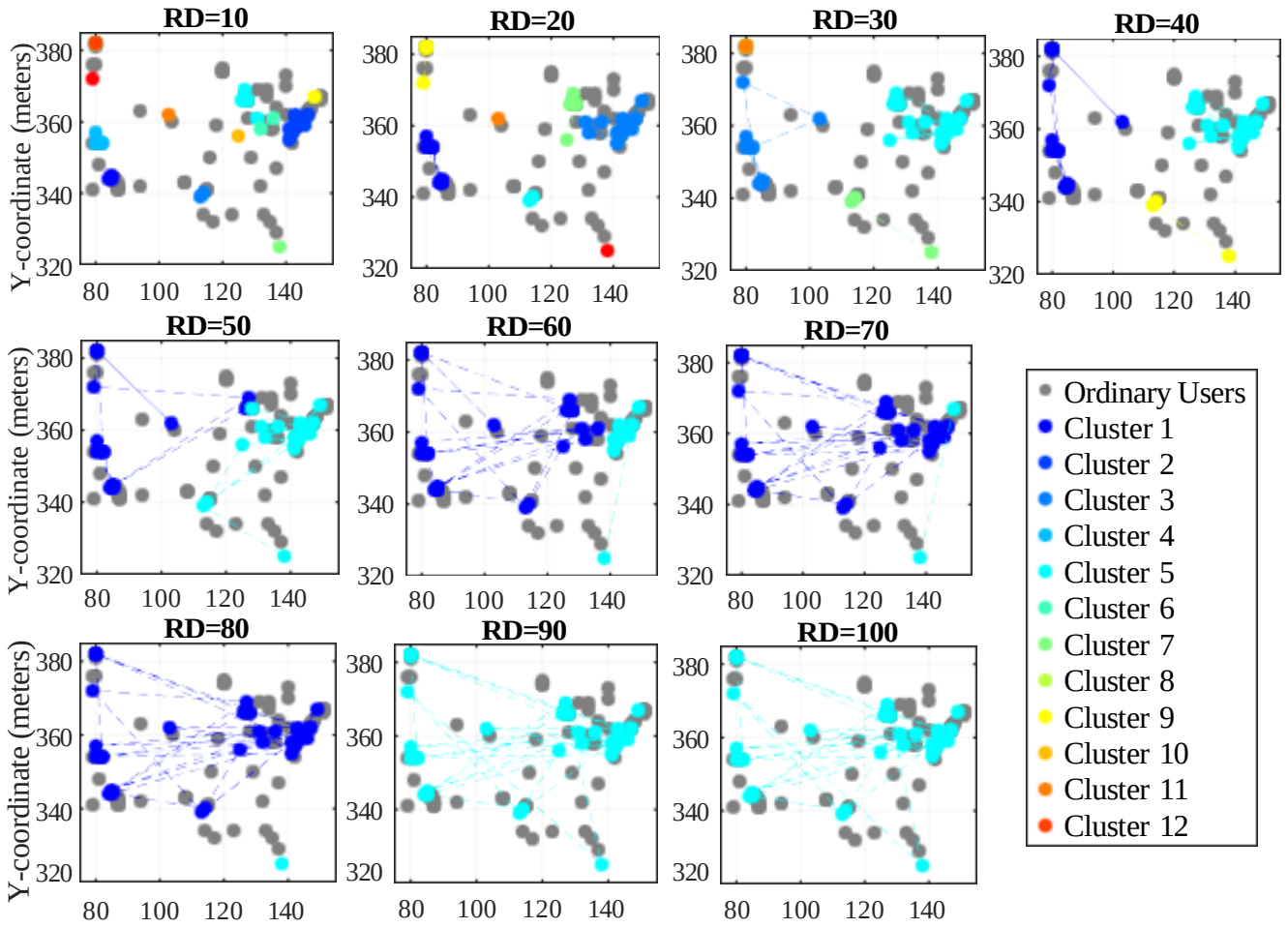

Figure 16. Clustering users for different $RD$ in scenario 7

# 6 CONCLUSIONS AND FUTURE WORKS

In this research, we have investigated the performance of a cellular network with D2D communication capabilities by proposing a method for predicting content popularity and utilizing the available storage space on users' devices. The proposed solution for the content popularity prediction problem has been examined using the XGboost machine learning model and cache placement.

Using a dataset, the behavior of the proposed approach has been evaluated in comparison to previous approaches under various scenarios, such as different cache sizes. The simulation results show that the proposed approach is capable of providing optimal performance under various network parameter conditions and can meet the research objectives, namely maximizing backhaul link offloading and increasing cache efficiency while considering limitations such as cache size, the number of helper users, and the maximum D2D communication range. Furthermore, a comparison of the proposed approach with previous methods shows that this approach provides a better hit rate and cache efficiency under various conditions. Additionally, it has an acceptable execution time in terms of practicality. To improve the efficiency and development of the proposed method in this research, the following points can be considered:

1. Diversity in cache size: Allow for changes in the cache size of helper devices and considering variable states during execution.

2. Combined caching: Utilizing storage space in base stations along with user devices to improve the proposed method.

3. Incentive mechanisms for participation: Investigating monetary, reputation-based, and priority-access incentives to encourage helper participation and enhance cache efficiency.

4. Advanced interference management: Studying dynamic frequency allocation, adaptive power control, and beamforming to further improve D2D interference mitigation and spectral efficiency.

5. Scalable clustering: Exploring scalable clustering methods—such as KD-tree spatial indexing and priority-based greedy selection—to improve performance in large-scale D2D networks.

6. Advanced Comparative Analysis: Extending the comparative study to include recent, sophisticated baselines, such as the DRL-based framework from [8], to assess trade-offs between efficiency and deep learning complexity.

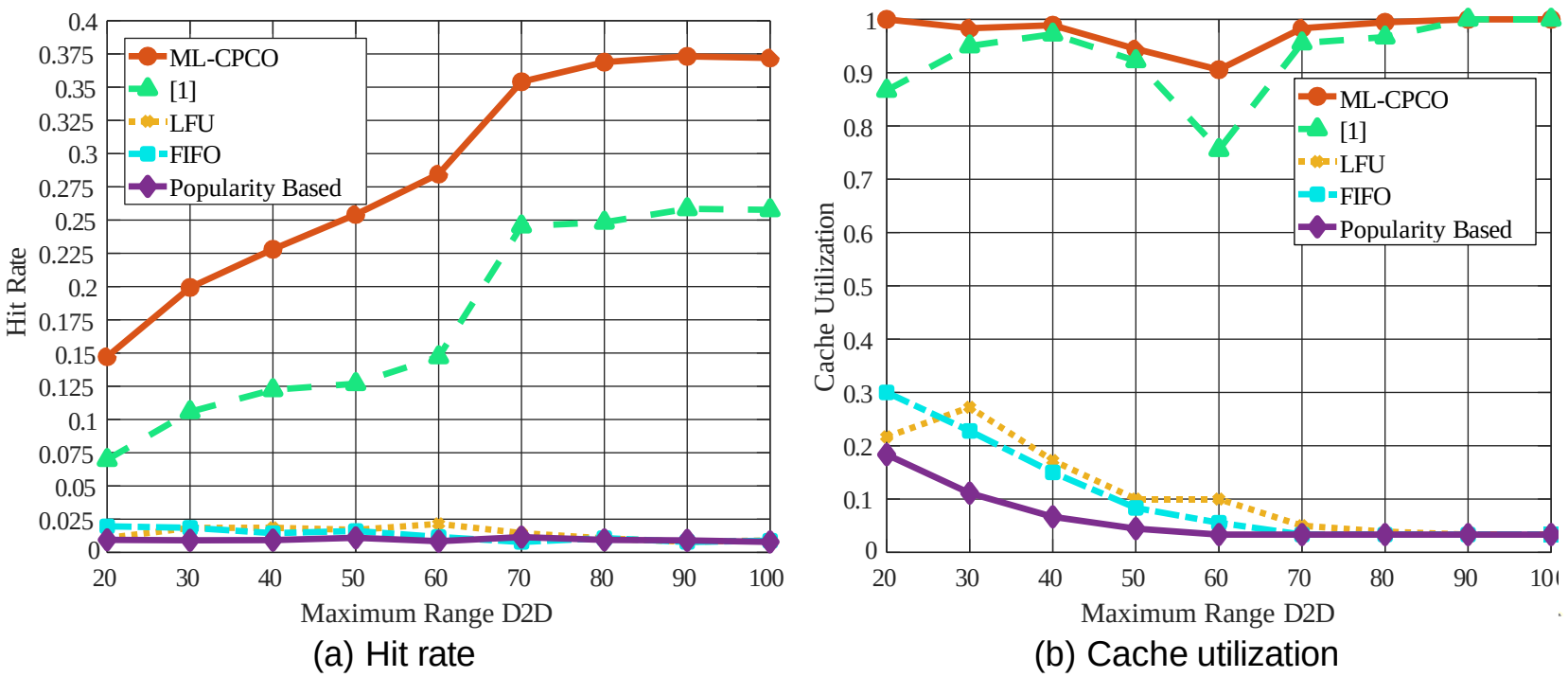

(a) Hit rate

(b) Cache utilization

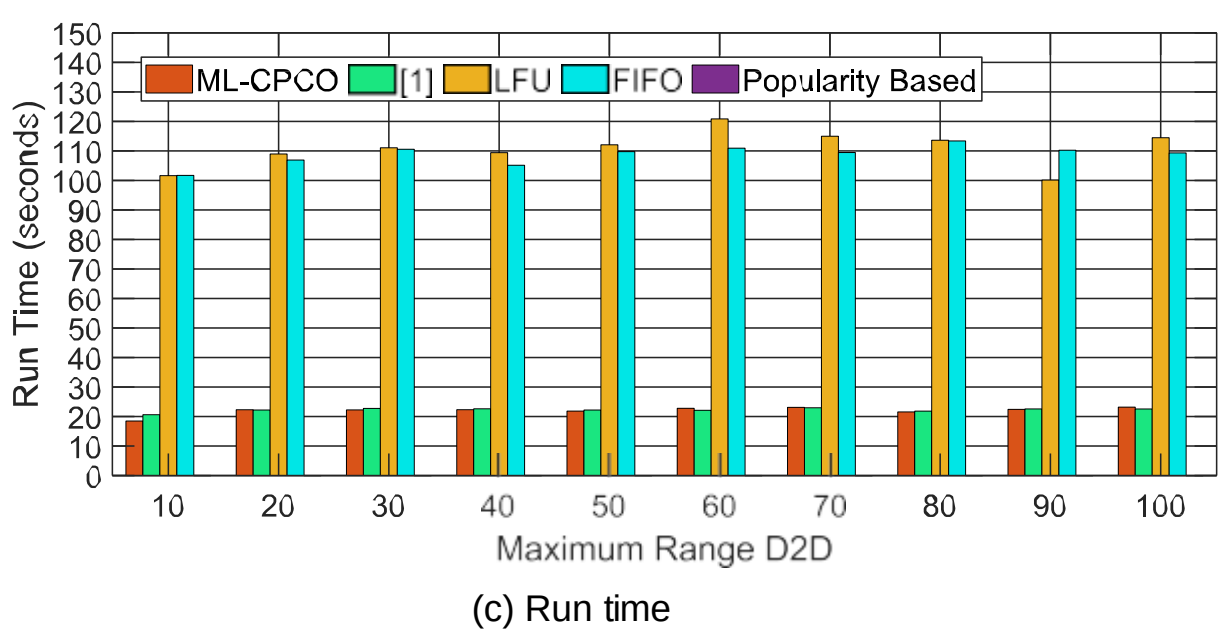

(c) Run time

Figure 17. Comparison of different evaluation criteria under changing RD in scenario 7

These suggestions can be considered as future approaches to developing and improving the performance of the proposed method.

## References

[1] A. Kazez and T. Girici, "Clustering-based device-to-device cache placement," *Ad Hoc Networks*, vol. 84, pp. 170–177, 2019.

[2] L. Li, G. Zhao, and R. Blum, "A survey of caching techniques in cellular networks: Research issues and challenges in content placement and delivery strategies," *IEEE Communications Surveys & Tutorials*, vol. 20, no. 3, pp. 1710–1732, 2018.

[3] H. Goian, O. Al-Jarrah, S. Muhaidat, Y. Al-Hammadi, P. Yoo, and M. Dianati, "Popularity-based video caching techniques for cache-enabled networks: A survey," *IEEE Access*, vol. 7, pp. 27 699–27 719, 2019.

[4] R. Amer, M. But, M. Bennis, and N. Marchetti, "Inter-cluster cooperation for wireless d2d caching networks," *IEEE Transactions on Wireless Communications*, vol. 17, no. 9, pp. 6108–6121, 2018.

[5] E. Bastug, M. Bennis, and M. Debbah, "Living on the edge: The role of proactive caching in 5g wireless networks," *IEEE Communications Magazine*, vol. 52, no. 8, pp. 82–89, 2014.

[6] Z. Hu, C. Fang, Z. Wang, S.-M. Tseng, and M. S. Obaidat, "Many-objective optimization-based content popularity prediction for cache-assisted cloud-edge-end collaborative iot networks," vol. 11, no. 1, 2024, pp. 1190–1200.

[7] P. Lin, Z. Ning, Z. Zhang, Y. Liu, F. R. Yu, and V. C. M. Leung, "Joint optimization of preference-aware caching and content migration in cost-efficient mobile edge networks," vol. 23, no. 5, 2024, pp. 4918–4931.

[8] M. Yan, M. Luo, C. A. Chan, A. F. Gygax, G. Chen, and C.-Y. Wang, "Energy-efficient content fetching strategies in cache-enabled d2d networks via an actor-critic reinforcement learning structure," vol. 73, no. 11, 2024, pp. 17 485–17 495.

[9] X. Zhang and J. Wang, "Heterogeneous statistical qos-driven resource allocation for d2d cluster-caching based 5g multimedia mobile wireless networks," in *2018 IEEE International Conference on Communications (ICC)*. IEEE, 2018, pp. 1–6.

[10] M. Ji, G. Caire, and A. Molisch, "Fundamental limits of caching in wireless d2d networks," *IEEE Transactions on Information Theory*, vol. 62, no. 2, pp. 849–869, 2016.

[11] L. Tang, Q. Huang, A. Puntambekar, Y. Vigfusson, W. Lloyd, and K. Li, "Popularity prediction of facebook videos for higher quality streaming," in *2017 USENIX Annual Technical Conference (USENIX ATC 17)*, 2017, pp. 111–123.

[12] M. Somesula, R. Rout, and D. Somayajulu, "Greedy cooperative cache placement for mobile edge networks with user preferences prediction and adaptive clustering," *Ad Hoc Networks*, vol. 140, p. 103051, 2023.

[13] R. Amer, H. ElSawy, J. Kibilda, M. Butt, and N. Marchetti, "Performance analysis and optimization of cache-assisted comp for clustered d2d networks," *IEEE Transactions on Mobile Computing*, vol. 21, no. 4, pp. 1334–1348, 2020.

[14] J. Wu, J. Zhang, and Y. Ji, "Dcec: D2d-enabled cost-aware cooperative caching in mec networks," *Electronics*, vol. 12, no. 9, p. 1974, 2023.

[15] Q. Li, Y. Zhang, A. Pandharipande, X. Ge, and J. Zhang, "D2d-assisted caching on truncated zipf distribution," *IEEE Access*, vol. 7, pp. 13 411–13 421, 2019.

[16] N. Abdolkhani, M. Eslami, J. Haghighat, and W. Hamouda, "Optimal caching policy for d2d assisted cellular networks with different cache size devices," *IEEE Access*, vol. 10, pp. 99 353–99 360, 2022.

[17] K. Khan, A. Naeem, and A. Jamalipour, "Incentive-based caching and communication in a clustered d2d network," *IEEE Internet of Things Journal*, vol. 9, no. 5, pp. 3313–3320, 2021.

[18] Y. Fu, L. Salaün, X. Yang, W. Wen, and T. Quek, "Caching efficiency maximization for device-to-device communication networks: A recommend to cache approach," *IEEE Transactions on Wireless Communications*, vol. 20, no. 10, pp. 6580–6594, 2021.

[19] Y. Qi, J. Luo, L. Gao, F.-C. Zheng, and L. Yu, "User preference and activity aware content sharing in wireless d2d caching networks," in *2020 IEEE/CIC International Conference on Communications in China (ICCC)*, 2020, pp. 987–992.

[20] M.-C. Lee and A. Molisch, "Individual preference aware caching policy design in wireless d2d networks," *IEEE Transactions on Wireless Communications*, vol. 19, no. 8, pp. 5589–5604, 2020.

[21] M. Yasir, S. Zaman, T. Maqsoo, F. Rehman, and S. Mustafa, "Copup: Content popularity and user preferences aware content caching framework in mobile edge computing," *Cluster Computing*, vol. 26, pp. 267–281, 2023.

[22] J. Bai, S. Zhu, Y. Ren, and P. Zhang, "The joint optimization of caching and content delivery in air-ground cooperation environment," 2025.

[23] M.-H. Yang, M.-C. Lee, and Y.-W. P. Hong, "Joint caching and recommendation optimization from network and user perspectives in wireless d2d networks," vol. 73, no. 2, 2025, pp. 1233–1247.

[24] K. Thar, N. Tran, T. Oo, and C. Hong, "Deepmec: Mobile edge caching using deep learning," *IEEE Access*, vol. 6, pp. 78 260–7827, 2018.

[25] A. Lekharu, P. Gupta, A. Sur, and M. Patra, "Collaborative video caching in the edge network using deep reinforcement learning," *ACM Transactions on Internet of Things*, 2024.

[26] Z. Sun and G. Chen, "Enhancing heterogeneous network performance: Advanced content popularity prediction and efficient caching," *Electronics*, vol. 13, no. 4, p. 794, 2025.

[27] M. Rezei and A. Montazerolghaem, "Software defined content popularity estimation for wireless d2d caching networks," in *2023 13th International Conference on Computer and Knowledge Engineering (ICCKE)*, 2023, pp. 426–431.

[28] F. M. Harper and J. A. Konstan, "The movielens datasets: History and context," *ACM Transactions on Interactive Intelligent Systems (TiiS)*, vol. 5, no. 4, pp. 19:1–19:19, 2015.

[29] Y. Jiang, M. Ma, M. Bennis, F.-C. Zheng, and X. You, "User preference learning-based edge caching for fog radio access network," *IEEE Transactions on Communication*, vol. 67, no. 2, pp. 1268–1283, 2018.

[30] S. Müller, O. Atan, M. van der Schaar, and A. Klein, "Context-aware proactive content caching with service differentiation in wireless networks," *IEEE Transactions on Wireless Communications*, vol. 16, no. 2, pp. 1024–1036, 2016.

[31] S. Li, J. Xu, M. van der Schaar, and W. Li, "Trend-aware video caching through online learning," *IEEE Transactions on Multimedia*, vol. 18, no. 12, pp. 2503–2516, 2016.